\pdfoutput=1
\documentclass[12pt]{article}
\usepackage{amssymb,amsmath,mathtools,mathrsfs,slashed}
\usepackage{color,graphicx}
\usepackage{cite}
\usepackage{amssymb,amsmath}
\usepackage{multirow}
\usepackage{array}
\usepackage{hyperref}
\usepackage{comment}
\usepackage{float}
\usepackage[top=3cm, bottom=3cm, left=2cm, right=2cm]{geometry}	

\usepackage{tabularx}
\newcolumntype{Y}{>{\centering\arraybackslash}X}
\usepackage{booktabs}
\AtBeginDocument{
    \heavyrulewidth=.08em
        \lightrulewidth=.05em
        \cmidrulewidth=.03em
        \belowrulesep=.65ex
        \belowbottomsep=0pt
        \aboverulesep=.4ex
        \abovetopsep=0pt
        \cmidrulesep=\doublerulesep
        \cmidrulekern=.5em
        \defaultaddspace=.5em
}
\usepackage{xcolor}

\newcommand\fverb{\setbox\pippobox=\hbox\bgroup\verb}
\newcommand\fverbdo{\egroup\medskip\noindent%
                              \fbox{\unhbox\pippobox}\ }
\newcommand\fverbit{\egroup\item[\fbox{\unhbox\pippobox}]}
\newbox\pippobox

\newcommand{\beq} {\begin{equation}}
\newcommand{\eeq} {\end{equation}}
\newcommand{\beqa} {\begin{eqnarray}}
\newcommand{\eeqa} {\end{eqnarray}}

\newcommand{\T}{\mathcal{T}}
\newcommand{\R}{\mathcal{R}}

\newcommand{\be}{\begin{equation}}
\newcommand{\ee}{\end{equation}}
\newcommand{\bea}{\begin{eqnarray}}
\newcommand{\eea}{\end{eqnarray}}

\def\cN{{\cal N}}

\newcommand{\vev}[1]{\left\langle#1\right\rangle}

\numberwithin{equation}{section}

\begin{document}
 
\begin{flushright}
HIP-2019-19/TH
\end{flushright}

\begin{center}

\centering{\Large {\bf Novel color superconducting phases of \\
$\cal{N}$ = 4 super Yang-Mills at strong coupling}}

\vspace{8mm}

\renewcommand\thefootnote{\mbox{$\fnsymbol{footnote}$}}
Oscar Henriksson,${}^{1,2}$\footnote{oscar.henriksson@helsinki.fi}
Carlos Hoyos,${}^{3,4}$\footnote{hoyoscarlos@uniovi.es} and
Niko Jokela${}^{1,2}$\footnote{niko.jokela@helsinki.fi}

\vspace{4mm}
${}^1${\small \sl Department of Physics} and ${}^2${\small \sl Helsinki Institute of Physics} \\
{\small \sl P.O.Box 64} \\
{\small \sl FIN-00014 University of Helsinki, Finland}

\vspace{2mm}
\vskip 0.2cm
${}^3${\small \sl Department of Physics, Universidad de Oviedo} \\
{\small \sl c/ Federico Garc\'{\i}a Lorca 18, 33007 Oviedo, Spain} 

\vspace{2mm}
\vskip 0.2cm
${}^4${\small \sl Instituto Universitario de Ciencias y Tecnolog\'{\i}as Espaciales de Asturias (ICTEA)}\\
{\small \sl  Calle de la Independencia, 13, 33004 Oviedo, Spain}

\end{center}

\vspace{8mm}

\setcounter{footnote}{0}
\renewcommand\thefootnote{\mbox{\arabic{footnote}}}

\begin{abstract}
\noindent 
We revisit the large-$N_c$ phase diagram of $\cal{N}$ = 4 super Yang-Mills theory at finite $R$-charge density and strong coupling, by means of the AdS/CFT correspondence. We conjecture new phases that result from a black hole shedding some of its charge through the nucleation of probe color D3-branes that remain at a finite distance from the black hole when the dual field theory lives on a sphere. In the corresponding ground states the color group is partially Higgsed, so these phases can be identified as having a type of color superconductivity. The new phases would appear at intermediate values of the $R$-charge chemical potential and we expect them to be metastable but long-lived in the large-$N_c$ limit.
\end{abstract}

\newpage
\tableofcontents


\newpage

\section{Introduction}\label{sec:introduction}

The original and one of the most studied examples of the AdS/CFT correspondence \cite{Maldacena:1997re,Gubser:1998bc,Witten:1998qj} builds on the $SU(N_c)$  $\cN=4$ super Yang-Mills theory ($\cN=4$ SYM). In the $N_c\to \infty$ and strong 't Hooft coupling limits $\cN=4$ SYM has been conjectured to be dual to type IIB superstring theory in $AdS_5\times S^5$. This geometry results as the near horizon limit of a stack of $N_c$ D3-branes. The conjecture has passed a number of non-trivial tests (see, \emph{e.g.}, \cite{Beisert:2010jr}) and it has also found applications 
in diverse fields \cite{CasalderreySolana:2011us,Ramallo:2013bua,Brambilla:2014jmp,Hartnoll:2016apf}.

One of the applications of the duality has been to use it as a playing ground for studying thermodynamics of strongly coupled gauge theories, starting with the seminal work of Witten \cite{Witten:1998zw}. In Witten's work, it was argued that when, {\emph{e.g.}}, the $\cN=4$ SYM theory is put on a spatial three-sphere, there will be a phase transition from a ``confined'' phase at low temperatures where the expectation value of the Polyakov loop vanishes to a ``deconfined'' phase at high temperatures where the expectation value is non-zero. Although the volume is finite, a phase transition is possible because $N_c\to \infty$ acts as a thermodynamic limit. In the gravity dual, the phase transition is manifested as the Hawking-Page transition \cite{Hawking:1982dh}, from empty global $AdS_5$ space to a black hole geometry. Soon thereafter, the analysis was extended to study the thermodynamic properties of states with $R$-charge dual to black holes with angular momentum along the $S^5$ directions \cite{Cai:1998ji,Cvetic:1999ne,Cvetic:1999rb,Chamblin:1999tk,Chamblin:1999hg}. In the $\cN=4$ SYM theory there are three independent $R$-charges corresponding to the rank of the global $R$-symmetry group $SU(4)_R$. Chemical potentials for each of the three charges can be introduced independently, although typically only a few symmetric cases have been considered, with charges that are either vanishing or equal to each other. In this paper we will give explicit results for the case of all chemical potentials equal ($\mu_1=\mu_2=\mu_3=\mu$) and for only one non-zero chemical potential ($\mu_1=\mu$, $\mu_2=\mu_3=0$), although most of our computations apply generally.

In \cite{Basu:2005pj,Yamada:2006rx,Hollowood:2008gp}, the results from the AdS/CFT calculation were compared with the phase diagram of $\cN=4$ SYM theory at weak coupling. The weak coupling analysis showed that above a critical value of the chemical potential $L\mu>1$ (in units of the radius of the three-sphere $L$), the theory does not have an obvious ground state, although there is a metastable deconfined phase that survives up to larger values of the chemical potential before becoming unstable. The value of $L\mu$ where the deconfined phase becomes unstable was seen to increase with the temperature. The analysis of \cite{Yamada:2007gb,Yamada:2008em} indicated that on the gravity side, black hole solutions at $L\mu>1$ were also metastable and will decay through the emission of D3-branes, a process which was dubbed ``brane fragmentation''. In the context of holographic applications to condensed matter physics this has also been called ``Fermi seasickness'' \cite{Hartnoll:2009ns}. In both the weak coupling and gravity dual description, the decay of the metastable state is exponentially suppressed $\sim e^{-N_c}$ in the large-$N_c$ limit. In the weak coupling calculation, there is an effective potential for the eigenvalues of $\cN=4$ SYM scalars that is unbounded from below but has a local minimum \cite{Hollowood:2008gp}. On the gravity side, D3-branes can lower their energy by escaping from the black hole, however this process is mitigated by a potential barrier between the region close to the horizon and the asymptotic region \cite{Yamada:2008em}. Although the gravity and field theory descriptions match qualitatively, it should be noted that the precise mechanism by which the black hole will lose its angular momentum has not been discussed in detail. We will not delve more into this issue, but assume that such mechanism exists and refer to it as ``brane nucleation''.

In this paper we will revisit the strong coupling phase diagram and in particular scrutinize the nature of metastable phases in the $L\mu>1$ region. In the original analysis of the nucleation instability, the effective potential for D3-branes was considered with fixed angular velocity. Because of this, the D3-brane experienced a centrifugal force that tended to expel it towards the boundary and which became the dominant effect for large enough rotation. Thus, it was observed that branes were expelled to the boundary as soon as some critical value of the velocity was reached. However, this would require that brane nucleation occurs in such a way that the angular momentum of the brane changes depending on the distance to the black hole horizon, eventually diverging as the asymptotic far region is reached. In this case the probe approximation should break down at asymptotically large distances from the horizon. 

We believe that this picture is not complete. Even in the grand canonical ensemble we expect the nucleation to be dominated by processes that conserve the total angular momentum of the black hole plus the emitted probe branes. The reasoning stems from the observation that the angular momentum is a conserved quantity for the D3-brane effective action, due to the isometries of the black hole geometry. A brane that nucleates from the stack will carry as much angular momentum as that lost by the black hole and will not affect to the asymptotic form of the geometry as long as it remains at a finite distance from the black hole. On the other hand, a process where the total angular momentum changes requires an exchange with the $AdS$ boundary and will imply a modification of the geometry throughout the whole space. Although both types of processes might be possible, the latter is likely to be more suppressed.

\begin{figure}[htb!]
   \begin{center}
   \includegraphics[scale=0.8]{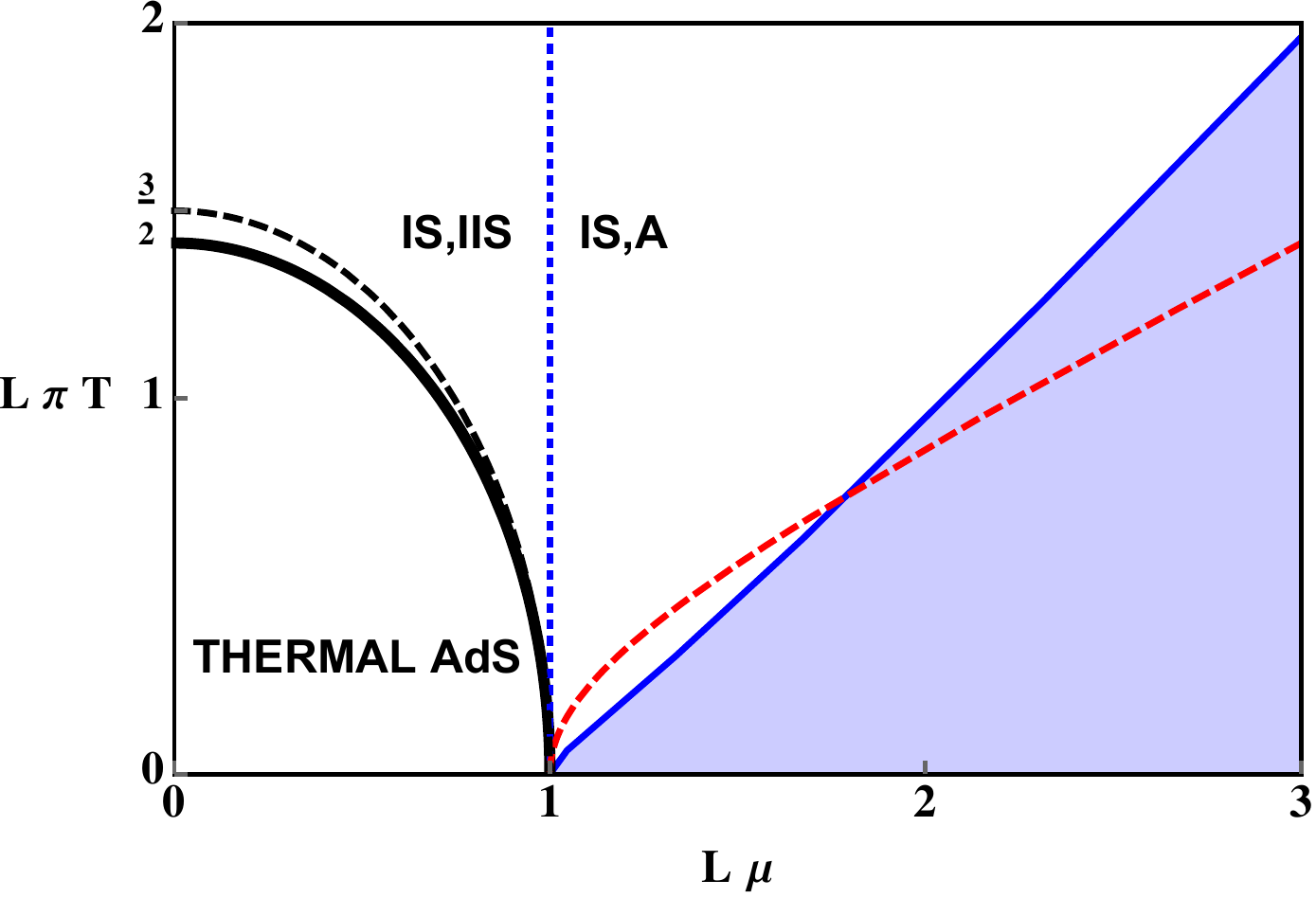}
     \caption{\small Phase diagram in the grand canonical ensemble for three equal chemical potentials. Susceptibilities diverge at the black thick curve, and black hole solutions exist only in the region outside the curve. The red dashed curve marks the onset of the instability for the symmetric phase ({\bf{IS}}) at higher values of the chemical potential. The blue dashed curve at $L\mu=1$ separates the unstable symmetric ({\bf{IIS}}) and asymmetric ({\bf{A}}) phases. The black dashed curve corresponds to the Hawking-Page transition.  The colored region shows where the black hole is unstable to brane nucleation. Outside that region and to the right of the $L\mu=1$ line the probe D3-brane effective potential has a minimum away from the horizon, but the nucleation instability is suppressed due to the mechanism described in the text (typical branes in the stack do not spin fast enough).}\label{fig:gcphd3}
   \end{center}
\end{figure}

Armed with these insights, we therefore expect the nucleation to happen first gradually through the emission of D3-branes with fixed angular momentum in such a way that the total angular momentum is preserved. This is a valid description as long as the timescales involved are short in comparison to those relevant for processes where the total angular momentum would change. Indeed, as we show in the body of the paper, the effective potential for D3-branes with fixed angular momentum is qualitatively different than the one for D3-branes with fixed angular velocity. In particular, when the $\cN=4$ SYM theory is on a sphere, the effective potential always increases at distances far away from the horizon thus preventing the nucleating branes from escaping to the asymptotic region. This allows for metastable configurations where a few D3-branes are localized at a finite distance from the black hole horizon. From the point of view of the field theory dual, this corresponds to a situation where some D3-branes are separated from the main stack. From the D3-brane effective theory vantage point, this is tantamount to having non-vanishing expectation values of some of the $\cN=4 $ SYM scalar fields. This Higgses the gauge group $SU(N_c)\rightarrow SU(N_c-n)\times U(n)$, for $n$ coincident branes localized outside the horizon, so there are new phases with spontaneously broken color symmetry.  It should be noted that this type of phase is not to be understood on par with the standard description of a color superconductor phase in QCD \cite{Rajagopal:2000wf}, which is described at weak coupling by a quark condensate and involves a locking between flavor and color symmetries. Nevertheless, in both cases there is a spontaneous breaking of the gauge group.

Our analysis also allows us to be more precise about the onset of this brane nucleation instability. While we recover the result that the instability is always present for $L\mu>1$, we argue that it will be greatly suppressed in a large part of this region of the phase diagram. The reason is that for large temperatures compared to the chemical potentials, a brane needs to have an exceedingly large angular momentum to nucleate in the bulk. However, we expect the $N_c$ spinning branes sourcing the background geometry to share the total angular momentum equally, with deviations from the average exponentially suppressed in the large-$N_c$ limit. For large $T/\mu$, a typical brane does not have enough angular momentum for it to nucleate, and thus the geometry is effectively stable. For the special case where the three chemical potentials are taken equal, this result is shown in the phase diagram of Fig.~\ref{fig:gcphd3}. There, stable charged black holes exists between the black and red dashed curves, corresponding to the Hawking-Page transition and the onset of thermodynamic instabilities, respectively. The colored blue region indicates where brane nucleation instability will {\em not} be suppressed by these statistical considerations. The full details of the phase diagram will be carefully explained in Secs.~\ref{sec:thermo} and \ref{sec:colorsc}.

In summary, we improve the analysis of the brane nucleation instability in backgrounds dual to finite density states of $\cN=4$ SYM on a sphere, and in the process find novel metastable color superconducting phases: the color superconducting matter is in a spatially homogeneous Higgs phase where the gluonic degrees of freedom obtained masses. We note that this is a prime and particularly clean example of a color superconducting phase in a top-down framework; for earlier interesting works in this context, see \cite{Chen:2009kx,Basu:2011yg,Rozali:2012ry,BitaghsirFadafan:2018iqr,Faedo:2018fjw}.

We have organized the paper as follows. In Sec.~\ref{sec:setup} we will review both the field theory and the spacetime geometry of spinning D3-branes. In Sec.~\ref{sec:thermo} we review the phase diagram focusing on the case with three equal chemical potentials. Then in Sec.~\ref{sec:D3probe} we will compute the effective potential for color probe D3-branes in the black hole geometry.  In Sec.~\ref{sec:colorsc} we discuss the nucleation instability and the color superconducting phases. In Sec.~\ref{sec:conclusions} we summarize and discuss future developments of our work. Finally, App.~\ref{app:thermo} contains derivation of the thermodynamic quantities through holographic renormalization, necessary details behind the conclusions in Sec.~\ref{sec:thermo}. In App.~\ref{app:singlechem} we discuss the case where only one chemical potential is nonzero.


\section{$R$-charged $\cN=4$ SYM}\label{sec:setup}

The $U(N_c)$ $\cN=4$  SYM theory is the low energy effective description of a stack of $N_c$ D3-branes in type IIB string theory. The $SO(6)$ symmetry of rotations in the space transverse to the D3-branes maps to the $SO(6)\cong SU(4)$ global $R$-symmetry acting on $\cN=4$ SYM fields. The AdS/CFT correspondence maps the (finite temperature) $SU(N_c)$ $\cN=4$ SYM theory to the near-horizon (black) $3$-brane geometry $\sim AdS_5\times S^5$, with $N_c$ units of five-form flux. The $R$-symmetry can be identified with the $SO(6)$ rotations of the $S^5$ component of the geometry, while the directions along the D3-brane worldvolume are embedded in the $AdS_5$ geometry. Following this map, introducing a nonzero $R$-charge in the $\cN=4$ SYM theory translates into a geometry with nonzero angular momentum along the $S^5$ directions.

\subsection{Field theory}

We follow the conventions of \cite{Yamada:2006rx,Hollowood:2008gp} in the parametrization of $\cN=4$ SYM fields and chemical potentials. The $\cN=4$ SYM theory involves a gauge field $A_\mu$, four Weyl fermions $\psi_i$, $i=1,\dots,4$, and six scalars $\phi_a$, $a=1,\dots,6$, all in the adjoint representation of the $SU(N_c)$ group. The gauge field is a singlet of the global $SU(4)_R$ symmetry, while the fermions and scalars furnish a fundamental $\mathbf{4}$ and an antisymmetric $\mathbf{6}$ irreducible representation, respectively. In the absence of chemical potentials or temperature, there is a moduli space that in an appropriate gauge is spanned by commuting constant values of the six scalars
\be
[\phi_a,\phi_b]=0\ ,\ \ \partial_\mu \phi_a=0\ , \ \forall\, a,b=1,\cdots,6 \ .
\ee
A basis can be chosen such that all the scalar fields are diagonal, so the moduli space is parametrized by the eigenvalues of the six scalars $\lambda_{a\alpha}$, $\alpha=1,\dots,N_c$. In the large-$N_c$ limit, one can usually apply a saddle-point approximation such that the ground state is characterized by an eigenvalue distribution. To each of the scalars one associates a direction in an $\mathbb{R}^6$ space. The distribution is defined as
\be
\rho(\vec{\lambda})=\sum_{\alpha=1}^{N_c} \delta^{(6)}(\vec{\lambda}-\vec{\lambda}_\alpha),\ \ \vec{\lambda}\in\mathbb{R}^6 \ .
\ee
Thus, the integral of the distribution over a region in $\mathbb{R}^6$ determines the number of eigenvalues of the six scalars that fall inside. In the large-$N_c$ limit the distribution can be approximated in many cases by a continuous function if the distance between the eigenvalues on $\mathbb{R}^6$ is small. In the dual gravity description, a continuous distribution determines the geometry (see e.g. \cite{Freedman:1999gk}), while a single isolated eigenvalue corresponds to a probe brane embedded in the geometry. In general, at nonzero temperature and/or chemical potential, the moduli space is lifted. Nevertheless, it is useful to consider the free energy for configurations where the scalar fields take constant values and to use eigenvalue distributions as a way to characterize the ground state in the large-$N_c$ limit.

The six scalars can be combined in three complex combinations $\Phi_A$, $A=1,2,3$,
\be
\Phi_1=\frac{1}{\sqrt{2}}(\phi_1+i\phi_2)\ ,\ \ \Phi_2=\frac{1}{\sqrt{2}}(\phi_3+i\phi_4)\ ,\ \ \Phi_3=\frac{1}{\sqrt{2}}(\phi_5+i\phi_6) \ .
\ee
Each of the fields is charged under a single component of the Cartan subalgebra, for which we introduce a chemical potential $\mu_1$, $\mu_2$, $\mu_3$. The fermions couple to a combination of these, see  \cite{Yamada:2006rx,Hollowood:2008gp} for details. Chemical potentials can be introduced as background fields in  the $\cN=4$ SYM action. We focus on the quadratic part of the scalar contribution, in Euclidean signature>:
\be
\begin{split}
&\mathcal L_{\phi^2}=\frac{1}{2g_{_{YM}}^2}{\rm tr}\left(\sum_{A=1}^3(D_\mu\phi_{2A-1}-i\mu_A \delta_{\mu,0}\phi_{2A})^2+\sum_{A=1}^3(D_\mu\phi_{2A}+i\mu_A \delta_{\mu,0}\phi_{2A-1})^2+\frac{1}{L^2}\sum_{a=1}^6\phi_a^2\right)+\ldots\\
&\mathcal L_{\phi^2}=\frac{1}{g_{_{YM}}^2}\sum_{A=1}^3 {\rm tr}\left(D_\mu \Phi_A^\dagger D_\mu\Phi_A+\left( \frac{1}{L^2} -\mu_A^2\right)\Phi_A^\dagger \Phi_A \right) +\sum_{A=1}^3 \mu_A J_A^0+\ldots 
\end{split}
\ee
The contribution of the scalars to the $R$-charge density can be read from the terms linear in the chemical potentials in the action
\be
J_A^0=\frac{i}{g_{_{YM}}^2}{\rm tr}\left( \Phi_A^\dagger \overset{\longleftrightarrow}{D_0} \Phi_A\right)+\text{fermions} \ .
\ee
The mass term is present due to the coupling of the scalars to the curvature of the spatial $S^3$ with radius $L$.  The chemical potentials add effectively a negative mass contribution, so in the absence of interactions the system is unstable when any of the chemical potentials becomes equal to $1/L$. Quantum corrections introduce additional terms in the effective potential, so that additional local minima may appear. In this case there can be metastable configurations where the eigenvalue distribution has support localized around those minima.

The terms depending on the chemical potential can be formally removed by factoring out a time-dependent phase of the scalar fields
\be
\Phi_A= e^{i\mu_A t} \widetilde{\Phi}_A \ .
\ee
Assuming $\tilde{\Phi}_A$ is time-independent, the $R$-charge becomes 
\be
J_A^0=-\frac{2\mu_A}{g_{_{YM}}^2} {\rm tr}\left(\widetilde{\Phi}_A^\dagger \widetilde{\Phi}_A \right) \ .
\ee
Neglecting for a moment the contribution to the effective potential due to the curvature of the $S^3$, the classical potential for the rescaled $\tilde{\Phi}_A$ fields is the same as the potential for the original set of scalars in the absence of a chemical potential.  Non-zero $R$-charge ground states would then be characterized by a distribution of rotating eigenvalues in $\mathbb{R}^6$, with angular velocities determined by the chemical potentials.

Note that in the large-$N_c$ limit, a continuous distribution of rotating eigenvalues can still be stationary if it is rotationally invariant around the origin of the moduli space. In this case, the dual geometry would then be stationary as well. On the other hand, isolated eigenvalues contributing to the $R$-charge density will enter as probe branes rotating in the internal directions associated to the $R$-symmetry. It is natural in this case to identify the angular velocities of the rotating branes in the internal space with the velocities of eigenvalues in the angular directions of $\mathbb{R}^6$. The distance to the horizon of the probe branes in the dual geometry should map to the distance from the origin of the moduli space of the isolated eigenvalues, at least qualitatively. Therefore, the effective potential for the probe branes that we will compute in Sec.~\ref{sec:D3probe} could be interpreted as the effective potential for the eigenvalues of the scalar fields.

\subsection{Dual geometry}

The rotating black brane solutions were introduced in \cite{Behrndt:1998jd,Cvetic:1999xp}; here we review their main characteristics. The metric in ten dimensions can be written as 
\begin{equation}\label{eq:10Dmetric}
  ds_{10}^2= \tilde\Delta^{1/2}ds_5^2 + R^2\tilde\Delta^{-1/2} \sum_{i=1}^3X_i^{-1} \left\{ d\sigma_i^2 + \sigma_i^2 \left( d\phi_i + R^{-1}A_i \right)^2 \right\} \ ,
\end{equation}
where $R$ is the AdS radius, $(R/l_s)^4 = 4\pi g_s N_c$, and
\begin{equation}
  \tilde\Delta = \sum_{i=1}^3 X_i \sigma_i^2 \ .
\end{equation}
The scalar fields $X_i$ satisfy $X_1 X_2 X_3=1$, and the three $\sigma_i$ satisfy $\sum_i \sigma_i^2=1$. The $\sigma_i$ can be parametrized with the angles on a two-sphere as
\begin{equation}
 \sigma_1 = \sin\theta \ , \qquad \sigma_2 = \cos\theta \sin\psi \ , \qquad \sigma_3 = \cos\theta \cos\psi \ .
\end{equation}

The 5D asymptotically-AdS metric $ds_5^2$ can be written as
\begin{equation}\label{eq:5Dmetric}
  ds_5^2 = - H(r)^{-2/3} f(r) dt^2
        + H(r)^{1/3} \big[ f(r)^{-1}dr^2 + r^2 d\Omega_{3,k}^2 \big] \ ,
\end{equation}
where
\begin{align}
  f(r) &= k - \frac{M}{r^2} + \left(\frac{r}{R}\right)^2H(r) \\
  H(r) &= H_1(r)H_2(r)H_3(r) \\
  H_i(r) &= 1 + \frac{q_i^2}{r^2} \ .
\end{align}
The coordinate $r$ is the usual holographic coordinate such that the boundary is at infinity. Here the parameter $k$ can take values 0, 1, or $-1$. This corresponds to the horizon geometry being $\mathbb{R}^3$, $S^3$, or $\mathbb{H}^3$, respectively. We are interested in the cases $k=0$ and $k=1$, and so we specialize to these values of $k$ from now on. In particular, we will consider the limiting case when the theory defined on a sphere approaches flat space. The unit metric $d\Omega_{3,k}^2$ is then
\begin{equation}\label{eq:3Dmetric}
 d\Omega_{3,k}^2 = \left\{ \begin{gathered} dx^2 + dy^2 + dz^2 \\ d\psi_{AdS}^2 + \sin^2\psi_{AdS}d\theta_{AdS}^2 + \sin^2\psi_{AdS}\sin^2\theta_{AdS}d\phi_{AdS}^2 \end{gathered} \ \begin{gathered} \text{for }k=0 \\ \text{for }k=1\ . \end{gathered} \right.
\end{equation}

We will adopt the conventions such that when $r\to \infty$ the metric approaches its canonical $AdS_5$ form:
\be
\begin{split}
k=0\ : \ \ &ds_5^2\simeq \frac{R^2}{r^2}dr^2+\frac{r^2}{R^2}\eta_{\mu\nu}dx^\mu dx^\nu \ \\
k=1\ : \ \ &ds_5^2\simeq \frac{R^2}{r^2}dr^2+\frac{r^2}{R^2}(-dt^2+L^2 d\Omega_3^2) \ ,
\end{split}
\ee
where $L$ is the radius of the sphere in the dual field theory. This limiting procedure needs to be supplemented with the following rescalings
\be\label{eq:restoreunits}
\begin{split}
k=0 \ : \ \ &(x,y,z)\to R^{-1}(x,y,z) \\
k=1 \ :\ \ &r\to (L/R) r\ \ ,\ \ t\to (R/L) t \ .
\end{split}
\ee
We will use these transformations when we compute thermodynamic variables in the dual theory, but for the moment we will continue using the original form of the metric \eqref{eq:5Dmetric}. 

The horizon radius $r_H$ is defined to be the largest root of $f(r)$. Using this condition we eliminate the parameter $M$ in favor of $r_H$:
\begin{equation}
 M = r_H^2\left(k + \frac{r_H^2}{R^2} H(r_H) \right) \ .
\end{equation}

The bulk scalars and gauge fields in these solutions are given by
\begin{equation}
 X_i(r) = H(r)^{1/3}/H_i(r)
\end{equation}
and
\begin{equation}\label{eq:gaugepots}
 A_i = \frac{q_i}{r_H^2+q_i^2}\sqrt{k (r_H^2 + q_i^2) + \frac{r_H^4}{R^2} H(r_H)} \left(1-\frac{r_H^2+q_i^2}{r^2+q_i^2}\right) dt \ .
\end{equation}
It is important that all the temporal gauge potentials vanish at the horizon, as in (\ref{eq:gaugepots}).

The self-dual 5-form field strength can be written in terms of the scalar field and 1-form gauge potentials as $G_5 + \star_{10} G_5$, with
\bea
 G_5  & = & (2/R)\sum_i\big(X_i^2\sigma_i^2 - \tilde\Delta X_i\big)\epsilon_5+(R/2) \sum_i (*_5\, d \ln X_i) \wedge d(\sigma_i^2) \nonumber\\
        & & +(R^2/2) \sum_i X_i^{-2}d(\sigma_i^2) \wedge\big( d\phi_i + R^{-1}A_i \big) \wedge *_5 F_i \ .
\eea
Here $F_i=dA_i$, $*_5$ denotes the Hodge dual with respect to the 5D metric $ds_5^2$, and $\epsilon_5$ is the corresponding 5D volume form. Using the solutions for the scalar and gauge fields we can find a 4-form potential $C_4$ such that $dC_4 = G_5$:
\bea
C_4 & = & \bigg[          \frac{r^4}{R} H(r)^{2/3}\tilde\Delta          - \frac{r_H^6}{R} \sum_i \frac{H(r_H)}{r_H^2+q_i^2}\, \sigma_i^2        \bigg] dt \wedge \epsilon_3 \nonumber \\
    &   & +\sum_i q_i\sqrt{k\, R^2 (r_H^2 + q_i^2) + r_H^4 H(r_H)}\ \sigma_i^2(R\, d\phi_i) \wedge \epsilon_3 \ .\label{eq:C4}
\eea
Here $\epsilon_3$ is the volume 3-form associated with $d\Omega_{3,k}^2$.


\section{Thermodynamics}\label{sec:thermo}

In this section we review the phase diagram of spinning D3-branes, that was studied previously in several works \cite{Cai:1998ji,Chamblin:1999tk,Cvetic:1999ne,Cvetic:1999rb}.
In general, the nucleation instability will only be relevant in the regions of the phase diagram where the classical supergravity description predicts a thermodynamically stable phase.  
The reader is therefore asked to pay attention to this regime when entering the later sections of this paper. To streamline the discussion, we have relegated the derivation of the thermodynamic quantities to App.\ref{app:thermo}. In the following, we present all the thermodynamic quantities in terms of parameters: 
\bea
{\rm flat\, space:}\  & x_i =\frac{q_i}{r_H}\ , \ & T_0=\frac{r_H}{\pi R^2}; \\
{\rm sphere:} \  & x_i =\frac{R}{L}\frac{q_i}{r_H}\ , \ & T_0=\frac{r_H}{\pi R^2}\ ,
\eea
where $L$ is the radius of the $S^3$. On the sphere, the temperature $T$, chemical potentials and charges $\mu_i$, $Q_i$, $i=1,2,3$, and the entropy density $s$ are
\bea 
 \mu_i & = & \pi T_0 x_i \left(\frac{\frac{1}{L^2 \pi^2 T_0^2}+\prod_{j\neq i}\left( 1+x_j^2\right)}{1+x_i^2}\right)^{1/2} \label{eq:mu} \\
 T     & = & T_0\left(1+\frac{1}{2L^2 \pi^2 T_0^2}+\frac{1}{2 }\sum_{i=1}^3 x_i^2-\frac{1}{2}\prod_{i=1}^3 x_i^2\right)\prod_{j=1}^3 \left(1+x_j^2\right)^{-1/2}   \label{eq:T} \\
 Q_i   & = & N_c^2\frac{\pi}{4}T_0^3 x_i\left(\frac{1+x_i^2}{L^2\pi^2T_0^2}+\prod_{j=1}^3\left( 1+x_j^2\right)\right)^{1/2}=\frac{N_c^2}{4}T_0^2\mu_i (1+x_i^2) \label{eq:Q} \\
 s &=& N_c^2 \frac{\pi^2}{2} T_0^3 \prod_{i=1}^3 \left(1+x_i^2\right)^{1/2}\ .  \label{eq:s}
\eea
The energy density is
\be
\varepsilon  =   \frac{3}{8} N_c^2 \pi^2 T_0^4\left(\prod_{i=1}^3 (1+x_i^2)+\frac{1}{L^2\pi^2 T_0^2}\left(1+\frac{2}{3}\sum_{i=1}^3 x_i^2\right)+\frac{1}{4L^4\pi^4 T_0^4}\right)\ . \label{eq:en}
\ee
The flat space values are recovered by sending $L\pi T_0\to \infty$ while keeping $x_i$ and $T_0$ fixed.

The energy density has a contribution that only depends on the radius of the sphere and that can be identified with the Casimir energy \cite{Balasubramanian:1999re}
\be
\varepsilon_{_\text{Casimir}}=\frac{3 N_c^2}{32 \pi^2}\frac{1}{L^4}\ .
\ee
After subtracting the Casimir energy contribution, the remainder can be identified with the internal energy density
\be
u= \frac{3}{8} N_c^2 \pi^2 T_0^4\left(\prod_{i=1}^3 (1+x_i^2)+\frac{1}{L^2\pi^2 T_0^2}\left(1+\frac{2}{3}\sum_{i=1}^3 x_i^2\right)\right)\ .
\ee
The energy density is a function of the entropy density and the charges. The temperature and chemical potential are obtained by taking thermodynamic derivatives of the internal energy. One can check that the following relations are satisfied by the expressions we have derived
\be
T=\frac{\partial u}{\partial s},\ \ \mu_i=\frac{\partial u}{\partial Q_i}\ .
\ee
The phase structure of the grand canonical ensemble is dictated by the Landau free energy density, which we will refer to as the grand canonical potential $\Omega$:
\be
\Omega=u-Ts -\sum_{i=1}^3 \mu_i Q_i\ .
\ee
The grand canonical potential is a function of temperature and the chemical potentials. In the flat space limit $L\pi T_0\to \infty$, the grand canonical potential is equal to minus the pressure and coincides with the renormalized on-shell action of the gravitational theory. However, the three quantities are in general different when they are computed on the sphere. 

We will not do a general analysis of thermodynamics but restrict to the case where the chemical potentials are all equal.\footnote{We examine the case where only a single chemical potential is nonzero in App.~\ref{app:singlechem}.}
Interestingly, there are two possible ways for the system to have the same total charge $Q=Q_1+Q_2+Q_3$ when $\mu_1=\mu_2=\mu_3=\mu$.
Either all the charges are equal $Q_1=Q_2=Q_3=Q/3$ with  $x_1=x_2=x_3=x$ or one of the charges is different from the other two $Q_1= Q_2\neq Q_3$, achieved by the following choice of normalized variables
\be\label{eq:xunequal}
x_1=x_2=x,\ \ x_3=\frac{1}{x L\pi T_0}\left(\frac{1+(L \pi T_0)^2(1+x^2)}{1+x^2}\right)^{1/2}\ .
\ee
We will refer to the phase with three equal charges as the symmetric phase and the phase with one unequal charge as the asymmetric phase. 
There are thus at least two different branches of solutions corresponding to two possible phases in the dual theory. The grand canonical potential is different in each case
\bea
 \Omega_{\bf{S}} & = & -N_c^2\frac{\pi^2}{8}T_0^4\left((1+x^2)^3-\frac{1}{L^2\pi^2 T_0^2}\right) \\
 \Omega_{\bf{A}} & = & -N_c^2\frac{\pi^2}{8}T_0^4\left((1+x^2)^3+\frac{1}{L^2\pi^2 T_0^2}\right)\ ,
\eea
where the subscripts of the potentials refer to ``symmetric'' and ``asymmetric'' phases, respectively. Note that the definition of the variable $x$ is different in each case, so equal values of $x$ do not correspond to equal values of temperature and chemical potential for the two different branches. The dominant branch will be the one with lower value of the potential at the same value of temperature and chemical potential.

\subsection{Phase diagram}

In the grand canonical ensemble we use as thermodynamic variables the dimensionless temperature and chemical potential, written in units of the sphere radius as $L\pi T$ and $L\mu$. It is convenient to parametrize their values in terms of $t_0=L\pi T_0\geq 0$ and $x\geq 0$ using the expressions in  \eqref{eq:mu}-\eqref{eq:en}.\footnote{Sending $x\to -x$ would flip the sign of the chemical potential and the charges, but the phase diagram is symmetric, so it is enough to consider positive values.}  The results are summarized in Fig.~\ref{fig:gcphd3}, where we also plot the results from the study of the nucleation instability that we discuss in the next sections. In the following we explain in more detail how it is derived.

To start with, we observe that there are several values of $(x,t_0)$ that yield the same values of $(L\mu,L\pi T)$, indicating several possible competing phases.  The phase with lower grand canonical potential will be the one dominating the thermodynamics, while other coexisting phases will then be either metastable or (thermodynamically) unstable. We will determine the thermodynamic stability from the matrix of susceptibilities
\be
\chi=L^2\frac{\partial\left(s,Q_1,Q_2,Q_3\right)}{\partial\left(T,\mu_1,\mu_2,\mu_3\right)}\Big|_{\mu_1=\mu_2=\mu_3=\mu}\ ,
\ee
where the $L^2$ factor is introduced to make a dimensionless combination. In order for the phase to be thermodynamically stable, all of the eigenvalues of the susceptibility matrix should be positive. The phase structure is summarized as follows.

\begin{itemize}
\item Symmetric phase ({\bf{S}}) \\
Black hole solutions exist outside the region around the origin of the $(L\mu,L\pi T)$ plane surrounded by the curve
\be\label{eq:limphd3}
(L\mu)^2+\frac{1}{2}(L\pi T)^2=1\ .
\ee
The are two branches of solutions that we label {\bf{IS}} and {\bf{IIS}}. {\bf{IIS}} is thermodynamically unstable and only exists for $L\mu \leq 1$. Notice that these solutions are continuously connected with the small Schwarzschild-black holes at $\mu=0$ and they play an interesting role in the microcanonical ensemble \cite{Asplund:2008xd,Jokela:2015sza,Hanada:2016pwv,Yaffe:2017axl,Berenstein:2018lrm,Hanada:2018zxn}. {\bf{IS}} is thermodynamically stable at lower values of the chemical potential but becomes unstable at the curve
\be\label{eq:instab3}
(L\mu)^2-4(L\pi T)^2=1\ .
\ee
The hyperbola approaches the straight line $L\mu=2L\pi T$, therefore for planar solutions, the symmetric phase becomes unstable for $\mu>2\pi T$. At the limiting curve \eqref{eq:limphd3} susceptibilities diverge to $\det \chi\to \pm \infty$, with the sign depending on the branch.

\item Asymmetric phase ({\bf{A}}) \\
Black hole solutions exist for $L\mu \geq 1$ and for any value of $L\pi T$. They are thermodynamically unstable in all their domain of existence. 
\end{itemize}

There are thermodynamically stable black hole solutions in the region of the phase diagram between the curves \eqref{eq:limphd3} and \eqref{eq:instab3}. Among the stable phases, there are two competing states, the black hole and global (thermal) $AdS_5$ without horizons. In order to determine which phase is dominant, we have to compare the value of the grand canonical potential for each phase. The global $AdS_5$ solution has $\Omega=0$ to leading order in the large-$N_c$ limit, so the black hole phase will be favored for $\Omega_{{\bf{S}}}<0$ and considered metastable otherwise.

\begin{figure}[t!]
   \begin{center}
\begin{tabular}{ccc}
   \includegraphics[width=0.31\textwidth]{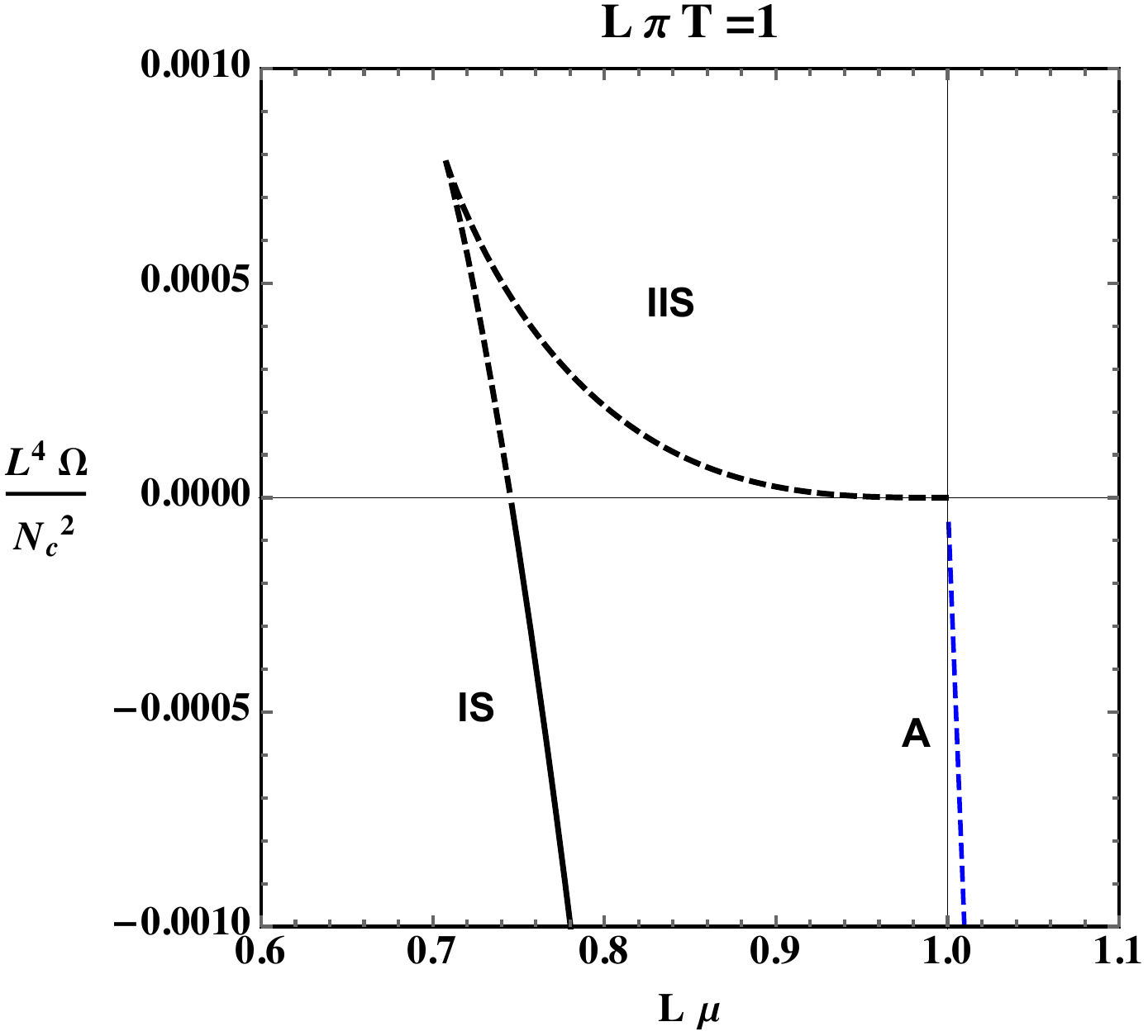} &
   \includegraphics[width=0.31\textwidth]{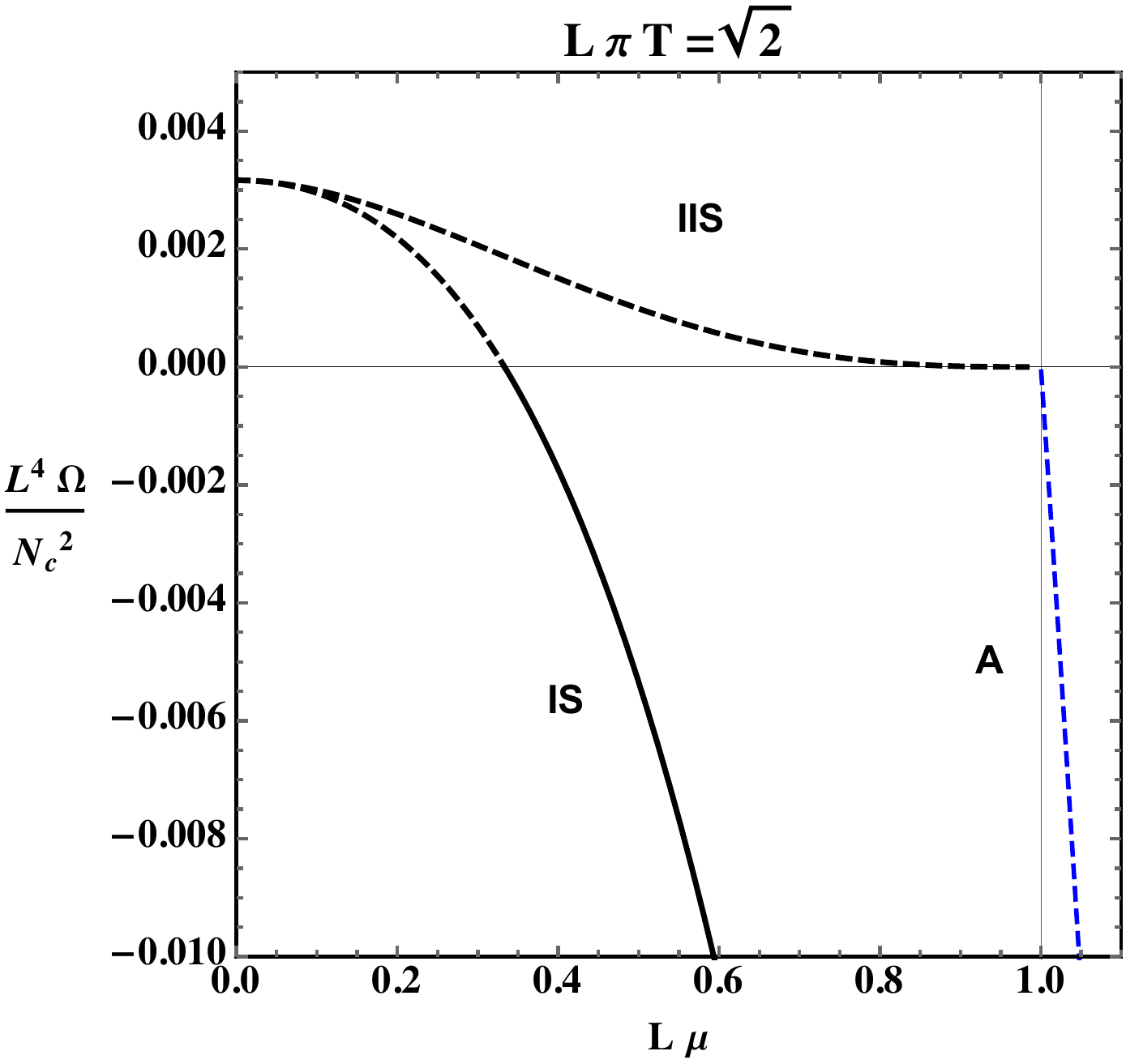} &
   \includegraphics[width=0.31\textwidth]{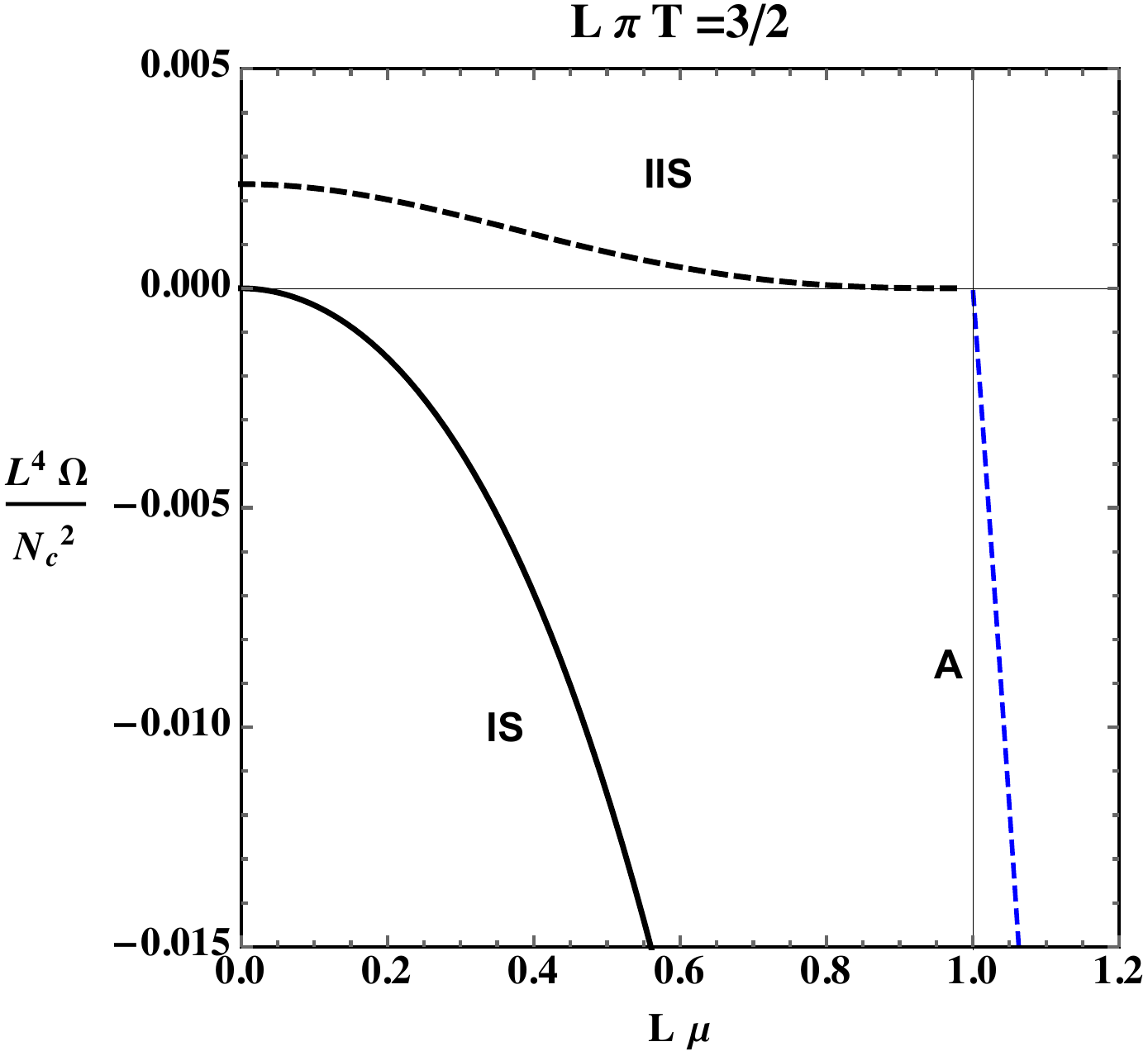}
   \end{tabular}
     \caption{\small Grand canonical potential as function of the chemical potential for different values of the temperature. Dashed lines correspond to unstable or metastable phases. The black curves are symmetric phases ({\bf{S}}) and the blue curve is for the asymmetric phase ({\bf{A}}). The symmetric branch with larger values of $\Omega$ corresponds to the unstable  phase ({\bf{IIS}}), while the lower branch ({\bf{IS}}) is metastable until it reaches $\Omega=0$, which corresponds to the Hawking-Page transition. Susceptibilities diverge at the point where the two symmetric branches touch. At high enough temperatures, $L\pi T>3/2$, the Hawking-Page transition goes away and the stable symmetric phase dominates at low values of the chemical potential. In all cases, at large enough values of the chemical potential, the symmetric phase becomes unstable when the asymmetric phase becomes dominant.}\label{fig:gcpot3}
   \end{center}
   \end{figure}

We have plotted the grand canonical potential as a function of the chemical potential at different values of the temperature in Fig.~\ref{fig:gcpot3}. We use dashed curves for the unstable or metastable phases. We see there that the limiting curves do not coincide with the Hawking-Page transition in general, which will be localized at the points where the grand canonical potential of the thermodynamically stable phase vanishes. For $\mu=0$ this happens at $L\pi T=\frac{3}{2}$, and above this temperature the stable black hole phase is always dominant. In  Fig.~\ref{fig:gcpot3}, the blue curve corresponding to the asymmetric phase eventually crosses the solid curve and the symmetric phase becomes unstable.

\section{Effective potential for probe color D3-branes}\label{sec:D3probe}

As we mentioned previously, the rotating geometry corresponds to the near horizon limit of a stack of spinning D3-branes whose low energy description is $\cN=4$ SYM with non-zero $R$-charge. In the absence of $R$-charge, and at zero temperature, there is a moduli space associated to the location of the D3-branes in the transverse space. If a single or few D3-branes are separated from the stack, this has a dual description as a single or few D3-branes inside the $AdS_5\times S^5$ geometry, that are treated as probes. The worldvolume of the D3-branes should be parallel to the ones forming the stack, so the branes are localized in the $S^5$ and radial $AdS_5$ directions and extended along the directions parallel to the $AdS_5$ boundary. For the case of global $AdS_5$ this actually means that the worldvolume wraps an $S^3$ and has finite volume. At non-zero $R$ charge and temperature, typically the moduli space is lifted, and probe branes in the dual geometry experience a force along the radial $AdS_5$ direction. We will now study the effective potential of probe branes in the holographic duals to $R$-charged states.

As we saw, the background solutions have three independent charges, corresponding to three mutually commuting $U(1)\subset SU(4)$ R-symmetries. From the 10D point of view, these charges can be viewed as angular momenta: the branes are spinning in the $S^5$. Probe branes in these backgrounds will be dragged along with the black hole rotation. Hence, we must allow them to rotate in (some of) the internal directions. The question of how to fix the angular velocity of the brane is then important. In \cite{Yamada:2008em} the probe brane was taken to spin with the angular velocity of the horizon, independently of the radial position. This, however, does not yield an accurate picture of the dynamics of the probe. A probe inserted at some radius will in general experience a force pushing it to larger or smaller radius. When this happens, the quantities that stay constant are not the angular velocities but the angular momentum of the probe. Thus, we will improve on the results of \cite{Yamada:2008em} by writing down expressions for the conserved quantities, which also include the energy of the probe. From this we construct the effective radial potential that the probe feels. This will allow us to make a more detailed analysis of possible instabilities.

We start from the action of a (probe) D3-brane,
\be\label{eq:D3action}
  S_{D3} =  -T_3\int d^4\xi \sqrt{-\det g_4}+T_3\int \hat C_4 \ ,
\ee
where $g_4$ is the pullback of the 10D metric (\ref{eq:10Dmetric}), $\hat C_4$ is the pullback of $C_4$ (\ref{eq:C4}) to the brane worldvolume, and $T_3=1/((2\pi)^3 g_s l_s^4)$.\footnote{We will use dimensionless worldvolume coordinates.} Recall that the dilaton is constant, so we omit it in the expression (\ref{eq:D3action}). We parametrize the timelike direction of the worldvolume of the brane by its proper time $\tau$, and denote the spacetime coordinates by capital letters $X^\mu(\tau)$. Allowing the brane to move in time, in the radial coordinate, and in the $\phi_i$ coordinates, the 10-velocity of the brane can be written as
\be
U \equiv \frac{dX^\mu}{d\tau}\partial_\mu = \dot \T(\tau)\partial_t + \dot \R(\tau)\partial_r + \sum_{i=1}^{3} \dot \Phi_i(\tau)\partial_{\phi_i} \ ,
\ee
where the dot denotes a derivative with respect to $\tau$. Since $\tau$ is the proper time, the 10-velocity squares to minus one,
\be\label{eq:10velocity}
 U_\mu U^\mu = -\dot \T^2 g_{tt} + \dot \R^2 g_{rr} + \sum_{i=1}^{3}\left( 2\dot \T\dot \Phi_i g_{t\phi_i} + \dot\Phi_i^2 g_{\phi_i\phi_i} \right) = -1 \ .
\ee
Here and below $g_{\mu\nu}$ denote components of the 10D metric (\ref{eq:10Dmetric}), with an extra minus sign in the definition of $g_{tt}$ such that it is positive. The induced metric on the brane worldvolume is
\be
 ds^2_4 = -\left[ \dot \T^2 g_{tt} - \dot \R^2 g_{rr} - \sum_{i=1}^{3}\left( 2\dot \T\dot \Phi_i g_{t\phi_i}+\dot\Phi_i^2 g_{\phi_i\phi_i} \right) \right] d\tau^2 + \sum_{i=1}^{3}g_{\chi_i \chi_i} d\chi_i^2 \ ,
\ee
where we call the spatial coordinates $\chi_i$. The pullback of $C_4$ becomes
\begin{equation}
\begin{split}
 \hat C_4 =& \left[ (C_4)_t \dot \T + \sum_{i=1}^{3} (C_4)_{\phi_i} \dot\Phi_i \right] d\tau\wedge \epsilon_3 \ ,
\end{split}
\end{equation}
where $(C_4)_t$ and $(C_4)_{\phi_i}$ are the $dt\wedge \epsilon_3$ and the $d\phi_i\wedge \epsilon_3$ components of $C_4$ in (\ref{eq:C4}), respectively. The action we find is thus
\bea
 S_{D3}  & = & -T_3\int d^4\xi \sqrt{-\det g_4}+T_3\int \hat C_4 \nonumber \\
   & = & -T_3 \int d^4\xi \Bigg\{ \sqrt{g_\Omega} \left[ \dot \T^2 g_{tt} - \dot \R^2 g_{rr} - \sum_{i=1}^{3}\left( 2\dot \T\dot \Phi_i g_{t\phi_i}+\dot\Phi_i^2 g_{\phi_i\phi_i} \right) \right]^{1/2} \nonumber\\
  & & - (C_4)_t\, \dot \T - \sum_{i=1}^{3} (C_4)_{\phi_i} \dot\Phi_i \Bigg\} \equiv \int d^4\xi\, \mathcal{L} \ .
\eea
The action does not depend on $\T$ or $\Phi_i$ explicitly, only their derivatives, making them cyclic variables. Thus, we can find the corresponding conserved energy and angular momentum densities using Noether's theorem:
\begin{align}
 E &\equiv -\frac{1}{T_3}\frac{\partial\mathcal{L}}{\partial \dot \T} = \frac{\sqrt{g_\Omega}\left(g_{tt}\dot \T - \sum_{i=1}^{3} g_{t\phi_i}\dot\Phi_i\right)}{\sqrt{ \dot \T^2 g_{tt} - \dot \R^2 g_{rr} - \sum_{i=1}^{3}\left( 2\dot \T\dot \Phi_i g_{t\phi_i}-\dot\Phi_i^2 g_{\phi_i\phi_i} \right) }} - (C_4)_t \notag \\
 &= \sqrt{g_\Omega}\left(g_{tt}\dot \T - \sum_{i=1}^{3} g_{t\phi_i} \dot\Phi_i \right) - (C_4)_t\label{eq:braneE} \\
 J_i &\equiv \frac{1}{T_3}\frac{\partial\mathcal{L}}{\partial \dot \Phi_i}= \frac{\sqrt{g_\Omega}\left(g_{t\phi_i}\dot \T + g_{\phi_i\phi_i}\dot\Phi_i\right)}{\sqrt{ \dot \T^2 g_{tt} - \dot \R^2 g_{rr} - \sum_{i=1}^{3}\left( 2\dot \T\dot \Phi_i g_{t\phi_i}-\dot\Phi_i^2 g_{\phi_i\phi_i} \right) }} + (C_4)_{\phi_i} \notag \\
 &= \sqrt{g_\Omega}\left(g_{t\phi_i}\dot \T + g_{\phi_i\phi_i}\dot\Phi_i\right) + (C_4)_{\phi_i}\label{eq:braneJ} \ .
\end{align}
We have simplified these expressions with the help of (\ref{eq:10velocity}). 

It is now possible to use (\ref{eq:10velocity}),(\ref{eq:braneE}), and (\ref{eq:braneJ}) to eliminate $\dot \T$ and $\dot\Phi_i$ and write an expression for the energy in terms of the angular momenta $J_i$ and background quantities:
\bea
 E & = & - (C_4)_t - \left( \sum_{i=1}^{3} g_{t\phi_i}\frac{J_i - (C_4)_{\phi_i}}{g_{\phi_i\phi_i}} \right) \nonumber \\
 & & + \sqrt{\left( g_{tt} + \sum_{i=1}^{3}\frac{g_{t\phi_i}^2}{g_{\phi_i\phi_i}} \right)\left(g_\Omega \left( 1+g_{rr}\dot \R^2 \right) + \sum_{i=1}^{3} \frac{\left( J_i - (C_4)_{\phi_i} \right)^2}{g_{\phi_i\phi_i}}  \right)} \ .
\eea
Note that since (\ref{eq:10velocity}) is quadratic in $\dot \T$ and $\dot \Phi_1$, there are two possible solutions; we have picked the one that has $\dot \T > 0$. Furthermore, note that we have reduced the system at hand to an effectively one-dimensional problem, depending only on the radial coordinate. As a last step, to get the effective potential that we are after, we set $\dot \R = 0$ in the previous expression $V_{\rm{eff}} \equiv E|_{\dot \R = 0}$:
\be
 V_{\rm{eff}}  = - (C_4)_t - \sum_{i=1}^{3} g_{t\phi_i}\frac{J_i - (C_4)_{\phi_i}}{g_{\phi_i\phi_i}} + \sqrt{\left( g_{tt} + \sum_{i=1}^{3}\frac{g_{t\phi_i}^2}{g_{\phi_i\phi_i}} \right)\left(g_\Omega + \sum_{i=1}^{3} \frac{\left( J_i - (C_4)_{\phi_i} \right)^2}{g_{\phi_i\phi_i}}  \right)} \ . \label{eq:Veff}
\ee
This is the general expression for the effective potential of a probe brane in this family of background solutions. In the following we analyze the effective potential separately for the spherical ($k=1$) and flat ($k=0$) horizon geometries given in \eqref{eq:3Dmetric}, corresponding to global AdS or the Poincar\'e patch, respectively.  Although the formulas for the potential are valid in general geometries, we will focus on the ensemble with three equal chemical potentials. As we reviewed in Sec.~\ref{sec:thermo}, the phase which is thermodynamically stable has three equal charges and we denoted it by {\bf{IS}}. Consequently, branes emitted by the black hole are expected to rotate with equal speeds in the three independent angles along the $S^5$. This lets us relate the three independent angular momenta in terms of one quantity, the total angular momentum $J_1+J_2+J_3=J_T$. With this choice the dependence on the angles of the $S^5$ drops out. In the case that the brane is wrapping a $S^3$ we additionally integrate over the volume to obtain the total energy. The total charge carried by the D3 brane is
\be
{\mathcal Q}_{\rm{D3}}=T_3 J_T.
\ee
For convenience, we will introduce $Q_{\rm{D3}}=R^{-3}{\mathcal Q}_{\rm{D3}}/N_c$ in units of charge density.


\paragraph{Global AdS}

\begin{figure}[h!]
   \begin{center}
   \includegraphics[scale=0.42,clip=true,trim=0pt 20pt 0pt 10pt]{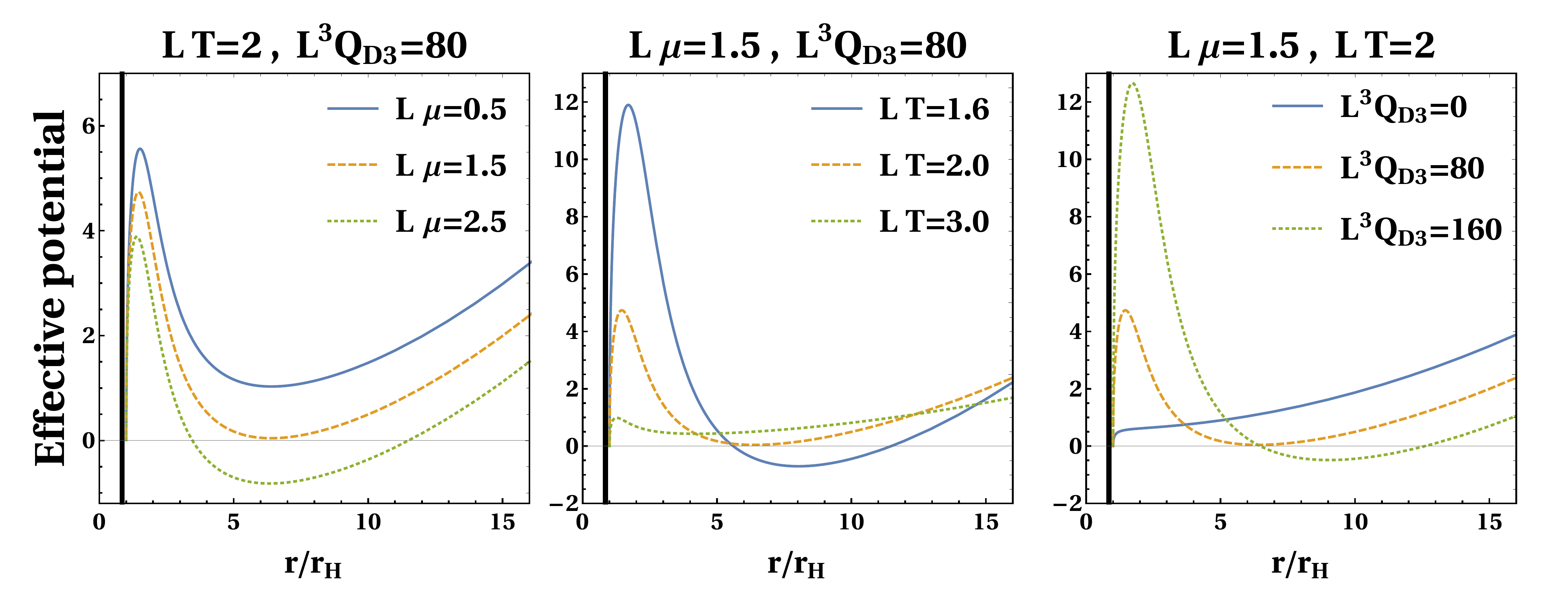}
   \caption{Probe brane effective potentials for $k=1$ and all charges equal, varying the chemical potential (left), the temperature (center), and the angular momentum of the probe (right).\label{fig:Veffk1QQQ}}
   \end{center}
\end{figure}

Setting $k=1$ in the black hole solutions we find that the effective potential always grows as $r^2$ for large radii (with the angular momenta held fixed). This can be understood from the fact that the radius of the $S^3$ that the D3-brane wraps grows with $r$. Since the D3-brane has a nonzero tension, increasing its volume translates in an increase in energy and eventually this growth becomes more important than the centrifugal force produced by the rotating geometry. It is interesting to compare this with the results of Yamada \cite{Yamada:2008em} where, as already mentioned, the probe was taken to rotate uniformly, with the same angular velocity as the horizon irrespective of the radial position. In that paper it was found that the coefficient of the $r^2$ term changed to negative for chemical potentials above some critical value $\mu_c$ (independent of temperature), implying that branes would escape to the AdS boundary. We argue that one should instead consider differentially rotating probes, in which case we find that the instability takes a somewhat different form.

It is convenient to introduce the dimensionless effective potential $\mathcal{V}_{eff} \equiv R\, V_{eff}/r_H^4$ and angular momentum $\mathcal{J}_T \equiv J_T/(r_H^3 R)=2 Q_{\rm{D3}}/(\pi T_0^3)$, as well as the similarly dimensionless quantities $t_0\equiv L\pi T_0$, $x=Q/r_H$, and $\rho \equiv r/r_H$, and then do the rescaling (\ref{eq:restoreunits}). We then arrive at the expression
\bea
  \mathcal{V}_{eff} & = & \frac{1-\rho^2}{x^2+\rho^2} \left(d\, \mathcal{J}_T-t_0^{-2}x^2-x^6+\left(1+3x^2\right)\rho^2+\rho^4\right)  \nonumber\\
  & &+\frac{1}{x^2+\rho^2}\bigg[ \left(1-\rho^2\right) \left(x^6 - \left(1+3x^2\right)\rho^2 - \rho^4 - t_0^{-2}\rho^2\right) \nonumber\\
  & &\times\Big( \left(t_0^{-2}x^2-2\, d\, \mathcal{J}_T\right)\left(1+x^2\right) + \mathcal{J}_T^2 + \left(x^2+\rho^2\right)^3 + x^2\left(1+x^2\right)^3 \Big) \bigg]^{1/2} \ , \label{eq:VeffDL}
\eea
where we have introduced the shorthand
\begin{equation}
 d \equiv x \sqrt{\frac{t_0^{-2}+(1+x^2)^2}{1+x^2}} \ .
\end{equation}
In Fig.~\ref{fig:Veffk1QQQ} we plot this dimensionless effective potential for the {\bf IS} phase in global AdS, as we vary the chemical potential, the temperature, and the probe brane angular momentum. We note that the potential always goes to zero (from above) at the horizon --- this is true in general, and signals the correct gauge choice for the four-form potential. At large radii, we notice the aforementioned $r^2$ growth. In between there is typically a local minimum. When this minimum dips below zero, it becomes a global minimum for the effective potential, and the system is then susceptible to the brane nucleation instability.

\begin{figure}[ht!]
   \begin{center}
   \includegraphics[width=0.45\textwidth]{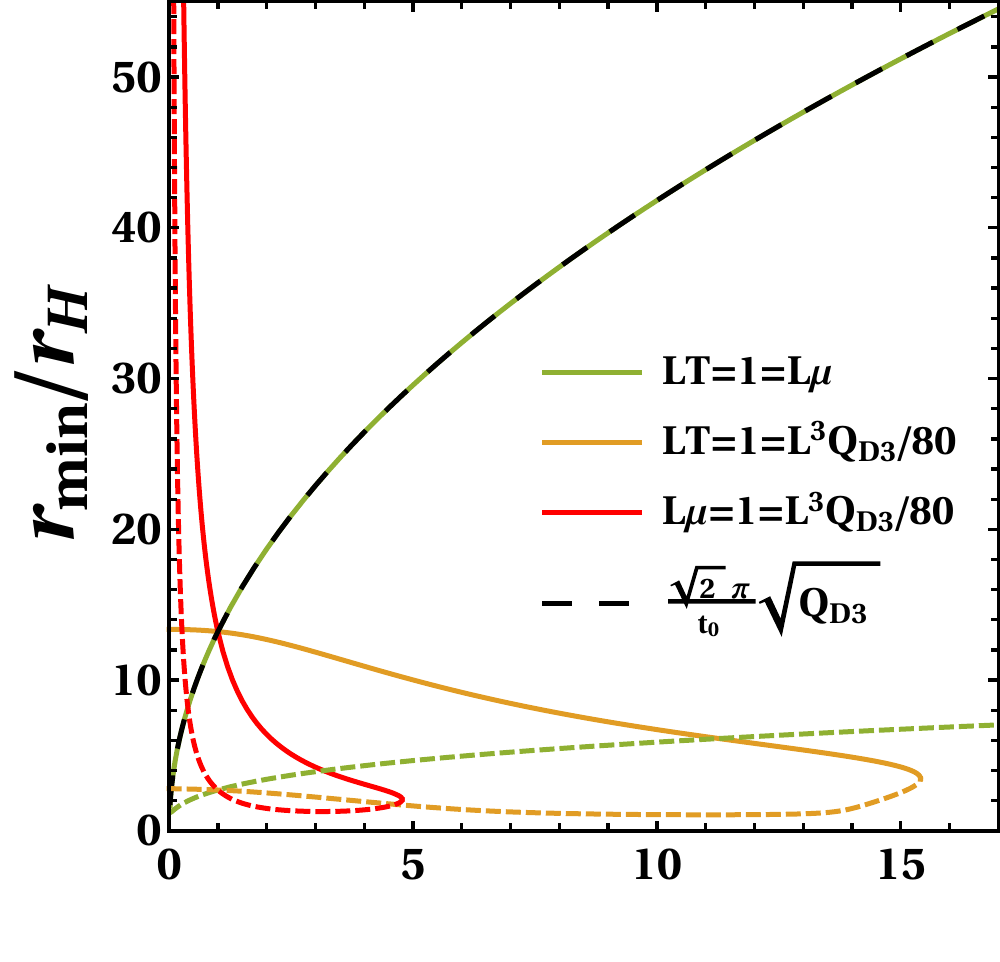}
   \caption{The minimum of the effective potential for $k=1$ as a function of $LT$ (red), $L\mu$ (orange), and $L^3 Q_{\rm{D3}}/80$ (green), when the other two values are kept fixed to unity. The solid parts of the curves correspond to minima of the effective potential, whereas dashed ones are the maxima. In all the cases the minimum radial coordinate is at a finite distance away from the horizon as there is always a potential barrier for the brane nucleation. Notice also that we have included the analytic result (\ref{eq:largeQD3}) in the limit of large $Q_{\rm{D3}}$ as the dashed black curve.}\label{fig:potmin}
   \end{center}
\end{figure}

In Fig.~\ref{fig:potmin} we have studied how the minimum of the effective potential moves as parameters $\mu,T,Q_{\rm{D3}}$ of the theory are varied. We learn that the minimum is always at a finite radial coordinate in the bulk. This is another manifestation of the fact that there is a potential barrier above the horizon for brane nucleation. The potential minimum is pushed asymptotically towards the boundary when $Q_{\rm{D3}}$ or $T^{-1}$ is increased without bound along curves where the other two parameters are kept fixed. Interestingly, the minimum is approaching the horizon as the chemical potential is increased. 

Interestingly, one can find an analytic expression for the minimum of the potential in the limit of large angular momentum of the brane. To obtain this, we use the fact that the minimum resides at large values of the radial coordinate and expand the effective potential for $\rho\to\infty$. Solving for the derivative $\partial_\rho\mathcal{V}_{eff}|_{\rho\to\infty}=0$, in the limit of large $L^3Q_{\rm{D3}}$, yields the remarkably simple result
\be\label{eq:largeQD3}
 \rho_{min} = \frac{\sqrt 2\pi}{t_0}\sqrt{Q_{\rm{D3}}}+O(1/\sqrt{Q_{\rm{D3}}}) \ , \ \ \ LT,L\mu = \rm{fixed} \ .
\ee
We have depicted this curve in the same Fig.~\ref{fig:potmin} with numerical data and it is spot on.

Remarkably, in the limit where the black hole carries a very large charge, it is also possible to find a simple analytic expression for the location of the minimum of the potential. The large charge limit in the symmetric case can be reached by taking
\be
L\pi T_0=t_0\to \infty, \ \ x_1=x_2=x_3=x=\sqrt{2}\left(1-\frac{\pi}{2\sqrt{3}}\frac{\tau}{t_0}\right) \ ,
\ee
with the parameter $\tau$ kept fixed, in the equations \eqref{eq:Q} and \eqref{eq:T}. The chemical potential \eqref{eq:mu} is large as well.  More precisely, to leading order the temperature and chemical potential are
\be
L  T\simeq  \tau +O(1/t_0), \ \ L\mu \simeq \sqrt{6} t_0+O(1)\ .
\ee
The charge density of the black hole becomes 
\be
\frac{L^3 Q}{N_c^2}\simeq \frac{9}{2\pi^2}\sqrt{\frac{3}{2}} t_0^3+O(t_0^2) \ .
\ee
We are interested in configurations where the charge of the D3-brane is of the order of $Q/N_c$. We can achieve this by taking $Q_{\rm{D3}}\sim L^3 Q/N_c^2$. A convenient normalization is
\be\label{eq:QD3fixed}
Q_{D3}=\frac{9}{4\pi^2}\sqrt{\frac{3}{2}} t_0^3 \chi_3\ .
\ee
\begin{figure}[ht!]
   \begin{center}
   \includegraphics[width=0.45\textwidth]{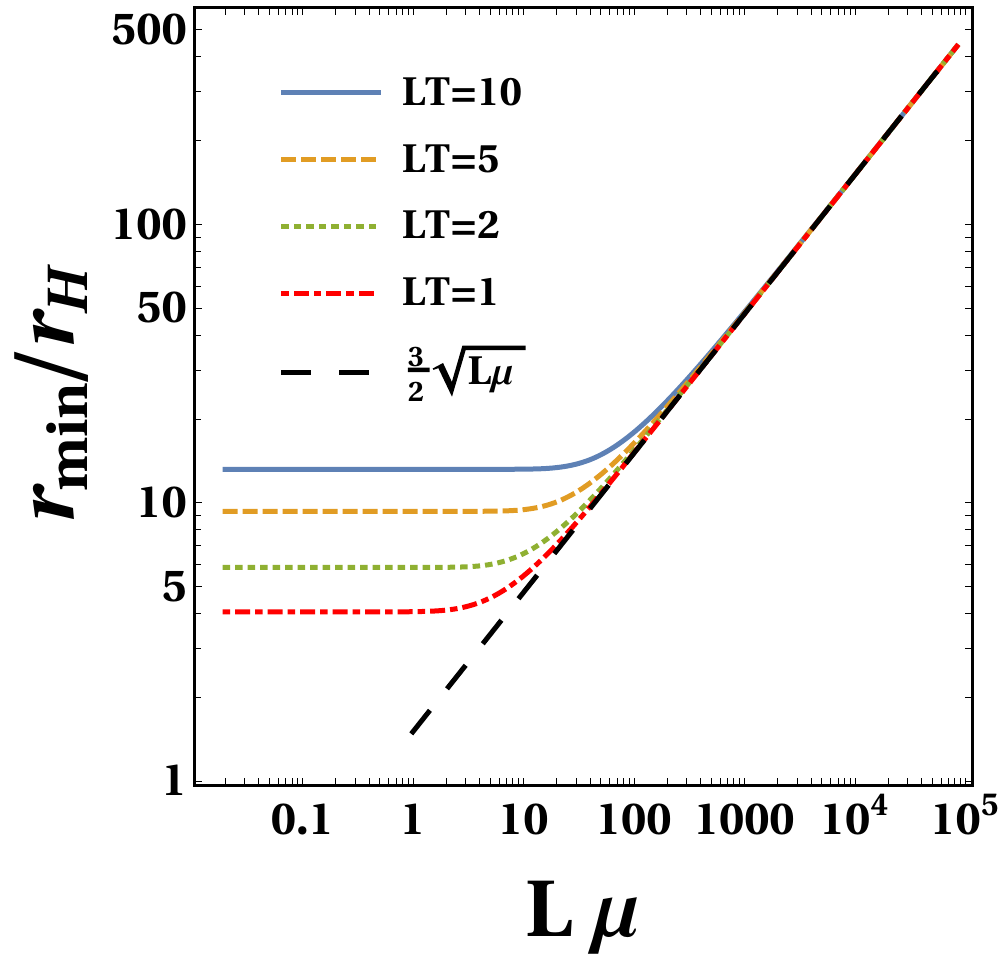}
   \caption{The minimum of the effective potential at fixed temperature and with the D3-brane charge fixed to $Q_{\rm{D3}}=\frac{9}{4\pi^2}\sqrt\frac{3}{2}t_0^3\chi_3$ as in (\ref{eq:QD3fixed}). We have chosen to present the curves for $\chi_3=1$ and $LT=10,5,2,1$ (top-down). The dashed black line is the analytic result $\frac{r_{min}}{r_H}\simeq\frac{3}{2}\sqrt{L\mu}$ (see (\ref{eq:gap})) for the gap in the large $R$-charge limit. Notice that the gap goes to a finite value in the opposite limit of small $R$-charge, in precise agreement with solving (\ref{eq:smallgap}) numerically.}\label{fig:largegap}
   \end{center}
\end{figure}

We evaluate the potential \eqref{eq:VeffDL}, rescale the radial coordinate as $\rho= t_0 u$ and expand for $t_0\to \infty$. We find
\be
  \mathcal{V}_{eff}  \simeq \frac{u^2}{2}+\frac{243 }{16 }\frac{\chi_3^2}{t_0^2}\frac{1}{u^2}+V_0\ ,
\ee
where the constant piece is
\be
V_0\simeq \frac{27}{2}\left(1-\chi_3\right)+O(\tau/t_0,1/t_0^2)\ .
\ee
The potential has a minimum at
\be
u_{min}=\left(\frac{3^5}{2^3}\right)^{1/4}\sqrt{\frac{\chi_3}{t_0}}\simeq 3\sqrt{\frac{3}{2}} \sqrt{\frac{\chi_3}{L\mu}} \ .
\ee
The  $\tau/t_0$ corrections to the potential only contribute to the constant part, so the location of the minimum is independent of the temperature in this limit. The value of the potential at the minimum is
\be
  \mathcal{V}_{eff}(u_{min})=\frac{27}{2}\left(1-\chi_3\right)+O(1/t_0,\tau/t_0) \ .
\ee
We see that the minimum of the potential is below its value at the horizon for
\be
\chi_3 > 1+O(1/t_0,\tau/t_0) \ .
\ee 
Therefore, $\chi_3 \sim 1$ and the location of the minimum has the following dependence with the chemical potential or the charge
\be\label{eq:gap}
\rho_{min}=t_0 u_{min} \simeq \frac{3}{2}\sqrt{L\mu}\sim (L^3Q/N_c^2)^{1/6} \ .
\ee

In the opposite limit of small $R$-charge, the chemical potential is small, and $x\approx 0$. In this case the effective potential reads
\be\label{eq:smallgap}
 \mathcal{V}_{eff}(\rho) =  1-\rho^4+\frac{1}{\rho t_0}\sqrt{\rho^2-1}\sqrt{\left(1+t_0^2(1+\rho^2)\right)\left(\rho^6+\frac{243\chi_3^2}{8}\right)}  + O(x) \ .
\ee
One straightforwardly finds from solving the minimum from this potential that it is at a finite radius $\rho>1$ for any temperature. 
In Fig.~\ref{fig:largegap} we have depicted the minimum of the potential as a function of chemical potential for various temperatures. The numerical results match precisely the obtained analytic results presented above. 

In the dual field theory one expects $\rho_{min}$ to be proportional to the distance from the origin to a minimum of the effective potential for the eigenvalues of the scalar fields. It would be interesting to check if a field theory calculation will give a similar chemical potential dependences, in particular that of $\sim\sqrt{\mu}$ in the limit of large $R$-charge.

\begin{figure}[ht!]
   \begin{center}
   \includegraphics[width=0.6\textwidth]{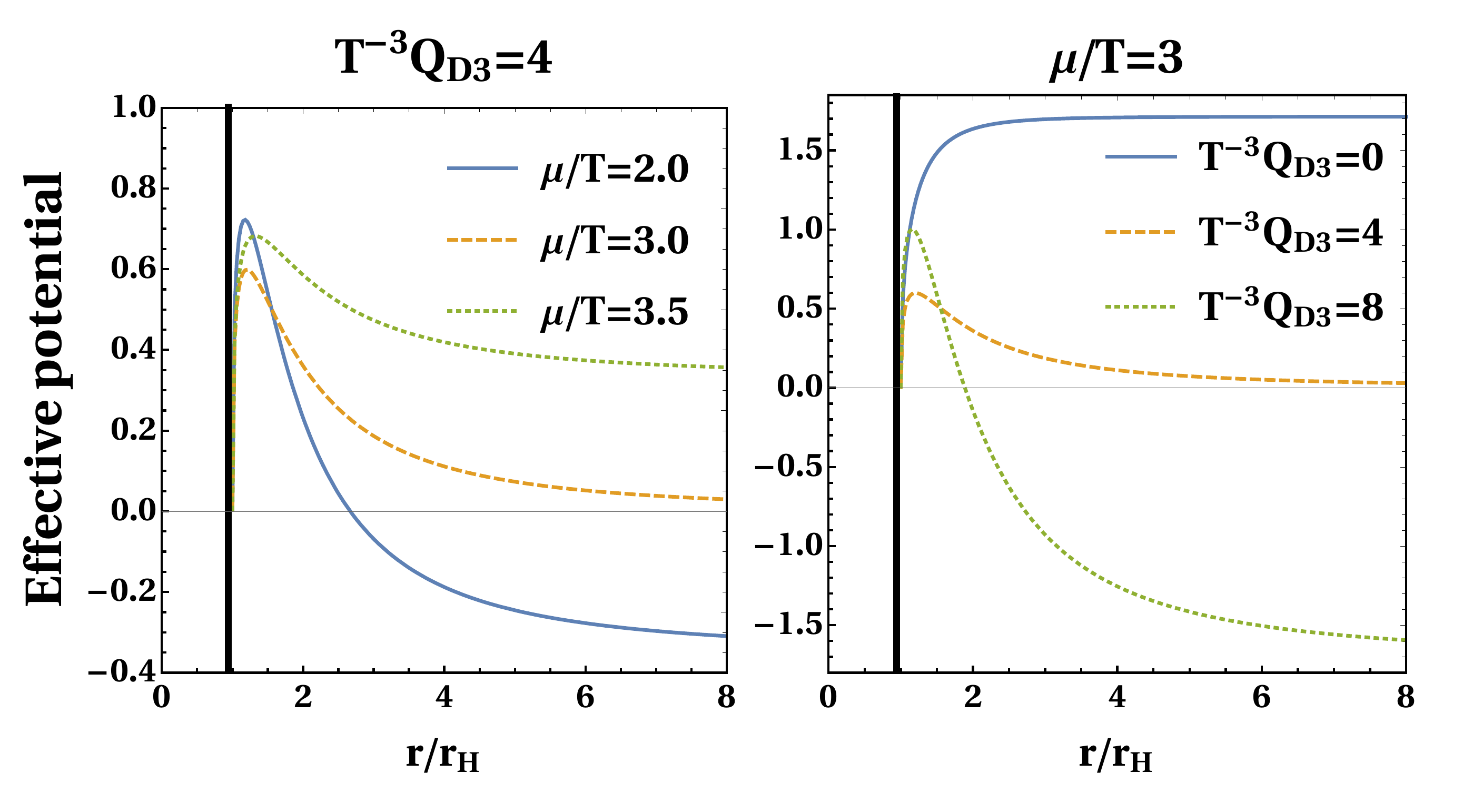}
   \caption{The effective potential for $k=0$ and all charges equal, varying $\mu/T$ (left) and the angular momentum of the probe (right).\label{fig:Veffk0QQQ}}
   \end{center}
\end{figure}

\paragraph{Poincar\'e patch}

It is also interesting to study the Poincar\'e patch of AdS, corresponding to putting the field theory in flat space. These results can be obtained easily by taking the limit $t_0\rightarrow\infty$ (corresponding to taking the 3-sphere radius $L$ large) in (\ref{eq:VeffDL}), or else by starting from (\ref{eq:Veff}) and setting $k=0$  in (\ref{eq:3Dmetric}). Fig.~\ref{fig:Veffk0QQQ} shows the effective potential in this case for the {\bf IS} phase, while varying $\mu/T$ and the probe angular momentum. While in global AdS the potential always grew as $r^2$ for large radii, in the Poincar\'e patch it always asymptotes to a constant value.

\section{Brane nucleation instability}\label{sec:colorsc}

The effective potential found in the previous section indicates that a probe brane placed at some radius will feel a force pulling it in towards the horizon or pushing it out towards infinity. The probe can minimize its energy by moving to the global minimum of the potential, which can be either at the black hole horizon, meaning the brane tends to fall into the hole, or at some point outside the horizon (including all the way out to the boundary).

Since the background is in fact sourced by a stack of D3-branes, the probe D3 effective potential can be used to detect instabilities. More precisely, a global minimum at the horizon signals that the geometry is stable; take a brane out of the stack and put it somewhere outside the horizon, and it wants to fall right back in. A global minimum outside the horizon, however, would imply that the geometry is only metastable; the system can minimize its energy by ``emitting'' branes from the black hole. 
Since these branes are the source of the $N_c$ units of flux corresponding to the rank of the gauge group, this type of instability would lead in the holographic dual to a spontaneous symmetry breaking of the gauge group  $SU(N_c)\to SU(N_c-1)\times U(1)$. When this happens in our nonzero density states, we interpret it as the onset of color superconductivity.

We find that the nucleation instability appears for $\mu>\mu_c=1/L$, in agreement with \cite{Yamada:2008em}. However, unlike the analysis at fixed angular velocity, the potential minimum is in this case at a finite distance from the horizon, instead of at infinity. Furthermore, this instability only occurs when the probe brane has a total angular momentum higher than a certain critical angular momentum $J_c(T,\mu)$. Above $J_c(T,\mu)$, the greater the angular momentum of the probe, the further away from the horizon is this minimum, and the more negative it is. Indeed, if we would allow the angular momentum of the probe to grow as a function of the radius as $r^2$, instead of being constant, we could reproduce the results of \cite{Yamada:2008em}. However, we expect that individual branes nucleating outside the horizon would have a fixed angular momentum, and would thus end up at some finite radius.

\begin{figure}[ht!]
 \begin{center}
  \includegraphics[width=0.6\textwidth]{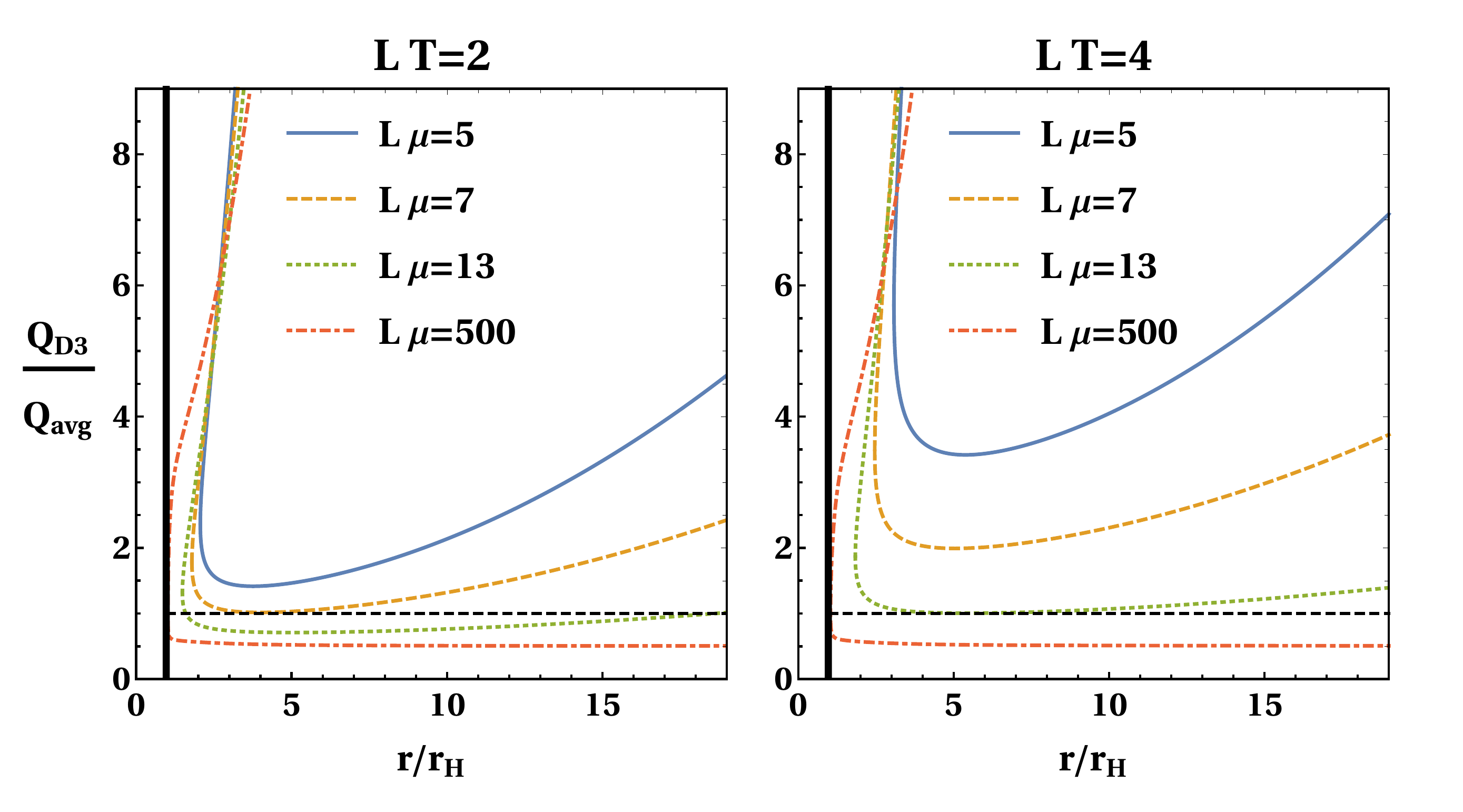}
  \caption{The curves show the angular momentum where the effective potential crosses zero as a function of radius, for backgrounds with $k=1$ and all charges equal.\label{fig:Jvsrk1QQQ}}
 \end{center}
\end{figure}

There are a few caveats to the discussion of the instability above. First, in general we find that when there is an instability, with a global minimum outside the horizon, there is also a potential barrier not far from the horizon. This is clearly seen in Fig.~\ref{fig:Veffk1QQQ}. If we want to think of the branes as being emitted from the horizon, then this would only be allowed quantum mechanically, and would thus be suppressed in the large-$N_c$ limit. We also note that the potential barrier in general grows with growing angular momentum, further suppressing the emission of large angular momentum branes. On the other hand, beyond the minimum there is a repulsive force, while anti-D3 branes would always feel an attractive force. This means that there could be an instability related to Schwinger pair production in the region between the maximum and the minimum of the potential, although there will also be some additional suppression related to the larger volume of branes further away from the horizon.

Second, since we are in the probe limit, we must remember to be careful when taking quantities such as the probe angular momenta large. If they are on the order of the total angular momenta of the background, the probe approximation will surely break down. Thus, if $J_c(T,\mu)$ is of order $N_c$, which can happen for very large temperatures $LT\sim N_c$, we can no longer go to our effective potential for guidance.

Third, nucleation of large angular momentum branes seems unlikely for the following reason: we could think of the nucleation as a semi-classical tunneling event, where a brane in the stack sourcing the background geometry takes a quantum leap from behind the horizon to region outside the horizon. There are $N_c$ of these branes, sharing a total angular momentum of order $N_c^2$. By standard statistical physics arguments we would expect the angular momentum to be evenly distributed over these branes, meaning each of them carries on average an angular momentum ${\mathcal Q}_{\rm{avg}}=Q/N_c\sim\mathcal{O}(N_c)$, with a relative variance that scales as $N_c^{-1/2}$. If the critical angular momentum density $Q_c=T_3 R^{-3} J_c(T,\mu)/N_c$ is larger than $Q_{\rm{avg}}=R^{-3}{\mathcal Q}_{\rm{avg}}/N_c$, the nucleation probability should be highly suppressed.

Taken together, this analysis suggests that the instability will be strongly suppressed when $N_c$ is large, and when the critical angular momentum $J_c(T,\mu)$ is larger than the average angular momentum of a brane in the stack, which should be ${\mathcal Q}_{\rm{avg}} = Q/N_c$. In Fig.~\ref{fig:Jvsrk1QQQ} we plot $Q_{\rm{D3}}/Q_{\rm{avg}}$, where $Q_{\rm{D3}}=T_3 R^{-3} J_T/N_c$ is proportional to the total angular momentum of the probe, as a function of radius for the all-charges-equal solution. The curves shown correspond to values of $Q_{\rm{D3}}/Q_{\rm{avg}}$ where the effective potential crosses zero, signaling a global minimum outside the horizon. If the curve stays above 1, we interpret it as saying that a typical brane in the stack does not have enough angular momentum to tunnel into the global minimum. The instability is therefore present, but strongly suppressed. When the curve dips below 1, which happens as we increase the chemical potential and/or lower the temperature, a typical brane does have enough angular momentum to prefer to sit at a global minimum in the bulk. Note that at large $N_c$ the tunneling is still suppressed, however. In Fig.~\ref{fig:gcphd3} we display the region of the phase diagram where the angular momentum for a typical brane of the stack is greater or equal to $Q_c$.

\begin{figure}[ht!]
 \begin{center}
  \includegraphics[width=0.4\textwidth]{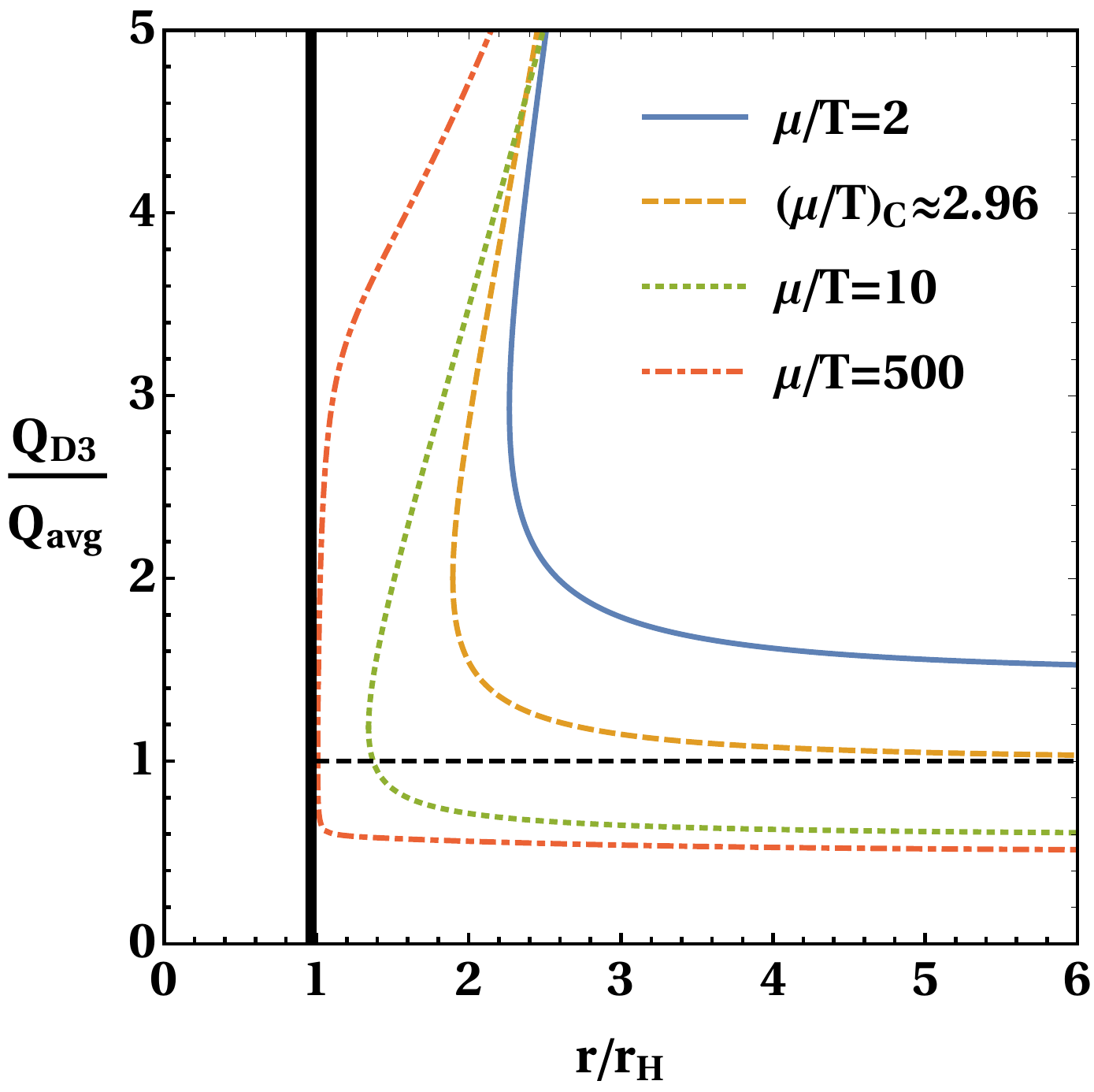}
  \caption{The curves show the angular momentum where the effective potential crosses zero as a function of radius, for backgrounds with $k=0$ and all charges equal. Each curve is at a fixed value of $\mu/T$.\label{fig:Jvsrk0QQQ}}
 \end{center}
\end{figure}

In the flat space limit  there are some qualitative changes relative to the finite volume case. The effective potential can now dip below zero for any value of the chemical potential or the temperature (or more precisely, for any value of the single dimensionless ratio $\mu/T$). The analysis in the previous subsections still applies, however. In particular, for small values of $\mu/T$, the probe needs angular momentum much larger than the expected average angular momentum $Q_{\rm{avg}}$ of a typical brane of the background color stack. Because of this, and the fact that also here the near-horizon potential barrier grows for large probe angular momentum, the instability is likely heavily suppressed in a large part of parameter space. More precisely, we find such suppression for $\mu/T>2\sqrt{2}\pi/3\approx 2.96$. In Fig.~\ref{fig:Jvsrk0QQQ}, we again plot the values of $Q_{\rm{D3}}/Q_{\rm{avg}}$ where the effective potential crosses zero as a function of radius.
   
Regarding the field theory interpretation of nucleation of probe branes, a possible picture could emerge from considering the expectation value of scalar fields of $\cN=4$ SYM. At zero temperature and chemical potential there is a moduli space for the scalars and vacuum states in the large-$N_c$ limit that can be characterized by a distribution of eigenvalues of the scalar fields on the moduli space. At nonzero temperature and chemical potential this description is still useful, even though most of the moduli space is lifted by an effective potential. In this case, stable or metastable states would be  characterized by an eigenvalue distribution that would be  localized around minima of the effective potential.
The initial black hole state would correspond to an eigenvalue distribution that is localized in a region around the origin of the moduli space.  Within states of the same charge, and at fixed chemical potential, there are other configurations where some weight of the distribution is taken from the region around the origin to another location further away (in the moduli space), in such a way that the free energy is lowered. These would be the new metastable phases described by probe branes outside the black hole. Eventually, there would be phase transitions where the total charge is changed and the weight of the eigenvalue distribution moves to asymptotically far regions where the effective potential is unbounded from below.

\section{Discussion and outlook}\label{sec:conclusions}

Due to the prominent role that it plays in digesting the AdS/CFT correspondence, the $SU(N_c)$ ${\cal{N}}=4$ supersymmetric Yang-Mills theory has been the focus of immense number of investigations. Fascinatingly, there are still secrets to be unlocked. In this paper, we have provided evidence of new phases of cold and dense matter that are expected at strong coupling.

An obvious objective of our endeavors is to construct the ground state for color superconducting matter.  This is both interesting and important in order to understand the phase structure of ${\cal{N}}=4$ SYM. However, the more pressing motivation is to make contact with phases of matter of QCD. To this end, in the near future we will report on our studies of a more realistic top-down holographic model, the so-called Klebanov-Witten model \cite{Klebanov:1998hh}. This model also shares the brane nucleation instability at low temperature in comparison to the baryon chemical potential \cite{Herzog:2009gd} and we plan to further explore the model by introducing an explicit breaking of conformal invariance by turning on masses to the hypermultiplets. This has the advantage that the model becomes increasingly closer to QCD and the phase diagram depending on ratios of both the temperature and the chemical potential to the new scale. We address where the color superconducting matter is the dominant homogeneous phase.

From our analysis it can be observed that the curvature of the spatial three-sphere acts as a stabilizing force for the probe branes in the bulk, impeding them from escaping to the $AdS$ boundary.  It would be interesting to find a similar stabilizing mechanism when the spatial directions are flat. This could potentially be achieved by attaching strings between the nucleating branes and the black hole horizon. They would be expected to give a contribution to the effective potential that grows with the separation between the probe brane and the black hole, thus potentially creating a global minimum at finite distance from the horizon.

Having established the true color superconducting ground state of a holographic model akin to QCD, there are many interesting repercussions to be followed. The regime of high density and small temperatures is very challenging for theoretical modeling, yet it is at the heart of contemporary high energy physics. The observation of gravitational waves of a coalescence of neutron stars \cite{TheLIGOScientific:2017qsa} has not only opened up a new observational window to astrophysics, but also enabled theorists to finding clues to pending questions on the behavior of dense matter. The key characteristic is the Equation of State which receives direct input through constraints on the mass-radius relationship, via tidal deformability, uncovered in GW170817. The ballpark estimates of neutron star bulk properties from holographic models have been highly successful \cite{Hoyos:2016zke,Hoyos:2016cob,Ecker:2017fyh,Annala:2017tqz,Jokela:2018ers,Ishii:2019gta,Chesler:2019osn}, so we are optimistic that this continues to be the case also for more exotic phases such as paired quark matter.

\medskip

\paragraph{Acknowledgments}

\noindent
We would like to thank Prem Kumar, Javier Tarr\'io, Aleksi Vuorinen, and Larry Yaffe for many useful discussions and comments on the draft version of this paper. O.~H. and N.~J. wish to thank Universidad de Oviedo for warm hospitality while this work was in progress. O.~H. is supported by the Academy of Finland grant no 1297472 and a grant from the Ruth and Nils-Erik Stenb\"ack foundation. C.~H. is supported by the Spanish grant MINECO-16-FPA2015-63667-P, the Ramon y Cajal fellowship RYC-2012-10370 and GRUPIN 18-174 research grant from Principado de Asturias. N.~J. has been supported in part by the Academy of Finland grant no. 1322307.

\appendix

\section{Derivation of thermodynamic quantities}\label{app:thermo}

We can extract the value of thermodynamic variables from the expansions of the 5D gauge field and metric at the boundary and the horizon. First we do the rescalings \eqref{eq:restoreunits} to fix the boundary metric to its correct form. The chemical potential is determined by the expansion of the gauge field at the boundary
\be
\mu_i=\frac{q_i}{R^2} \left(\frac{\prod_{j\neq i} \left(1+\frac{q_j^2}{r_H^2}\right)}{1+\frac{q_i^2}{r_H^2}}\right)^{1/2} \ .
\ee
The temperature is most easily extracted by performing a Wick rotation on the 5D metric to Euclidean signature and demanding that the geometry is smooth at the horizon
\be
T=\frac{r_H}{\pi R^2}\left(1+\frac{1}{2 }\sum_{i=1}^3 \frac{q_i^2}{r_H^2}-\frac{1}{2}\prod_{i=1}^3 \frac{q_i^2}{r_H^2}\right)\prod_{j=1}^3 \left(1+\frac{q_j^2}{r_H^2}\right)^{-1/2} \ .
\ee 

The charge densities can be computed using holographic renormalization
\be
Q_i=-\lim_{r\to\infty}\frac{R}{16\pi G_5}\sqrt{-g}g^{tt}g^{rr} \partial_r A_{i\, t}=\frac{R}{16\pi G_5}\frac{(H_1 H_2 H_3)^{2/3}}{X_i^2}\frac{r^3}{R^2} \partial_r A_{i\, t} \ .
\ee
In the last step we have used the fact that the electric flux remains constant along the radial coordinate. Introducing the explicit form of the solution in the equation above, the result for the charge densities is
\be
Q_i=\frac{q_i r_H^2}{8\pi G_5 R^3}\prod_{j=1}^3 \left(1+\frac{q_j^2}{r_H^2}\right)^{1/2} \ .
\ee
The overall factor can be expressed in terms of the radius of $AdS$ and the number of colors using the $AdS/CFT$ dictionary
\be
\frac{R^3}{G_5}=\frac{2 N_c^2}{\pi}\ .
\ee
Then, the charge densities are
\be
Q_i=\frac{N_c^2}{(2\pi)^2}\frac{q_i r_H^2}{R^6}\prod_{j=1}^3 \left(1+\frac{q_j^2}{r_H^2}\right)^{1/2}\ .
\ee
The entropy density is computed as the area of the black hole in Planck units, divided by the volume of the spatial directions along the boundary
\be
s=\frac{A_{BH}}{4 G_5 V_3}=\frac{N_c^2 }{2\pi}\frac{r_H^3}{R^6}\prod_{i=1}^3\left(1+\frac{q_i^2}{r_H^2}\right).
\ee

We can follow the same steps to compute the chemical potential, temperature, and charge densities in the case where the field theory lives on a sphere. The results are
\bea\label{eq:thermovar}
\mu_i & = & \frac{R}{L}\frac{q_i}{R^2}\left(\frac{\frac{R^4}{L^2 r_H^2}+\prod_{j\neq i}\left( 1+\frac{R^2}{L^2}\frac{q_j^2}{r_H^2}\right)}{1+\frac{R^2}{L^2}\frac{q_i^2}{r_H^2}}\right)^{1/2} \\
T & = & \frac{r_H}{\pi R^2}\left(1+\frac{R^4}{2L^2 r_H^2}+\frac{1}{2 }\frac{R^2}{L^2}\sum_{i=1}^3 \frac{q_i^2}{r_H^2}-\frac{1}{2}\prod_{i=1}^3 \frac{R^2}{L^2}\frac{q_i^2}{r_H^2}\right)\prod_{j=1}^3 \left(1+\frac{R^2}{L^2}\frac{q_j^2}{r_H^2}\right)^{-1/2} \\
Q_i & = & \frac{N_c^2}{(2\pi)^2}\frac{\frac{R}{L} q_i r_H^2}{R^6}\left(\frac{R^4}{L^2 r_H^2}+\prod_{j\neq i}\left( 1+\frac{R^2}{L^2}\frac{q_j^2}{r_H^2}\right)\right)^{1/2} \left( 1+\frac{R^2}{L^2}\frac{q_i^2}{r_H^2}\right)^{1/2} \\
s &= & \frac{N_c^2 }{2\pi}\frac{r_H^3}{R^6}\prod_{i=1}^3\left(1+\frac{R^2}{L^2}\frac{q_i^2}{r_H^2}\right)^{1/2} \ .
\eea

Let us define the variables
\bea
{\rm flat\, space:}\  & x_i =\frac{q_i}{r_H}\ , \ & T_0=\frac{r_H}{\pi R^2}; \\
{\rm sphere:} \  & x_i =\frac{R}{L}\frac{q_i}{r_H}\ , \ & T_0=\frac{r_H}{\pi R^2}\ ,
\eea
in terms of which the thermodynamic quantities take a somewhat simpler form. In flat space we have
\bea
 \mu_i & = & \pi T_0 x_i \left(\frac{\prod_{j\neq i} \left(1+x_j^2\right)}{1+x_i^2}\right)^{1/2} \label{eq:flatmu}\\
 T & = & T_0 \left(1+\frac{1}{2 }\sum_{i=1}^3x_i^2-\frac{1}{2}\prod_{i=1}^3 x_i^2\right)\prod_{j=1}^3 \left(1+x_j^2\right)^{-1/2} \label{eq:flatT}\\
 Q_i & = & N_c^2\frac{\pi}{4}T_0^3 x_i \prod_{j=1}^3 \left(1+x_j^2\right)^{1/2}=\frac{N_c^2}{4}T_0^2\mu_i (1+x_i^2) \label{eq:flatQ} \\
 s &=& N_c^2 \frac{\pi^2}{2} T_0^3 \prod_{i=1}^3 \left(1+x_i^2\right)^{1/2} \label{eq:flats} \ .
\eea
In the sphere the expressions are similar, but there is an additional dependence on the dimensionless combination $L\pi T_0$:
\bea
 \mu_i & = & \pi T_0 x_i \left(\frac{\frac{1}{L^2 \pi^2 T_0^2}+\prod_{j\neq i}\left( 1+x_j^2\right)}{1+x_i^2}\right)^{1/2} \\
 T     & = & T_0\left(1+\frac{1}{2L^2 \pi^2 T_0^2}+\frac{1}{2 }\sum_{i=1}^3 x_i^2-\frac{1}{2}\prod_{i=1}^3 x_i^2\right)\prod_{j=1}^3 \left(1+x_j^2\right)^{-1/2}  \\
 Q_i   & = & N_c^2\frac{\pi}{4}T_0^3 x_i\left(\frac{1+x_i^2}{L^2\pi^2T_0^2}+\prod_{j=1}^3\left( 1+x_j^2\right)\right)^{1/2}=\frac{N_c^2}{4}T_0^2\mu_i (1+x_i^2) \\
 s &=& N_c^2 \frac{\pi^2}{2} T_0^3 \prod_{i=1}^3 \left(1+x_i^2\right)^{1/2}\ .
\eea
The flat space values (\ref{eq:flatmu})-(\ref{eq:flats}) are recovered by sending $L\pi T_0\to \infty$ while keeping $x_i$ and $T_0$ fixed.

The energy density and pressure are determined by the expectation value of the energy-momentum tensor, which can be computed from the Brown-York tensor. For a radial slice, the  induced metric and extrinsic curvature are
\be
\gamma_{\mu\nu}=G_{\mu\nu}\ ,\ \ K_{\mu\nu}=\frac{1}{2 \sqrt{G_{rr}}} \partial_r G_{\mu\nu} \ .
\ee
The Brown-York tensor is 
\be
\pi_{BY}^{\mu\nu}= K^{\mu\nu}-\gamma^{\mu\nu} K \ ,
\ee
where the indices of the extrinsic curvature are raised with the induced metric and the trace is $K=\gamma^{\alpha\beta} K_{\alpha\beta}$. The expectation value of the energy momentum tensor is obtained from the $r\to \infty$ limit of the BY tensor with appropriate factors and the addition of boundary counterterms to cancel the divergent contributions \cite{Skenderis:2002wp}:
\be
\vev{T^{\mu\nu}}=\frac{1}{\sqrt{-h}}\lim_{r\to \infty} \frac{r^2}{R^2} \left[ -\frac{1}{8\pi G_5}\sqrt{-\gamma} \pi_{BY}^{\mu\nu}+\frac{\delta S_{ct}}{\delta \gamma^{\mu\nu}}\right] \ ,
\ee
where the boundary metric has been defined as
\be
h_{\mu\nu}=\lim_{r\to \infty} \frac{R^2}{r^2}\gamma_{\mu\nu}\ .
\ee
In this case to cancel the divergence we need just one counterterm, proportional to a boundary cosmological constant $S_{ct}\sim\int \sqrt{-\gamma}\Lambda$. However, for a generic solution of the gravitational action, the scalar fields will have a different behavior at the boundary, corresponding to turning on the non-normalizable modes. In the more general case additional counterterms proportional to masses for the scalar fields at the boundary are needed, and they should be kept even when the non-normalizable modes are turned off, as it is the case for the solutions we are studying. A counterterm action that contains both the cosmological constant and scalar mass terms is 
\be
S_{ct}=-\frac{1}{8\pi G_5} \int d^4 x \sqrt{-\gamma} \frac{2}{R}\sum_{i=1}^3 X_i^{-1}\ .
\ee
The result, for the planar black hole, is
\be
\vev{T^{00}}=\varepsilon \ ,\ \ \vev{T^{ij}}=p \delta^{ij}=\frac{\varepsilon}{3} \delta^{ij} \ ,
\ee
where the energy density is
\be
\varepsilon  =  \frac{3}{8} N_c^2 \pi^2 T_0^4\prod_{i=1}^3 (1+x_i^2)\,.
\ee
In the sphere the calculation is similar, but one needs to add a further counterterm proportional to the Ricci scalar of the induced metric
\be
S_{ct}=-\frac{1}{8\pi G_5} \int d^4 x \sqrt{-\gamma} \left(\frac{2}{R}\sum_{i=1}^3 X_i^{-1}-\frac{R}{2} {\cal R}[\gamma]\right)\ .
\ee
In this case the components of the energy-momentum tensor are
\be
\vev{T^{00}}=-\varepsilon h^{00}\ ,\ \ \vev{T^{ij}}=p h^{ij}=\frac{\varepsilon}{3} h^{ij} \ .
\ee
The energy density takes the form 
\be
\varepsilon  =  \frac{3}{8} N_c^2 \pi^2 T_0^4\left(\prod_{i=1}^3 (1+x_i^2)+\frac{1}{L^2\pi^2 T_0^2}\left(1+\frac{2}{3}\sum_{i=1}^3 x_i^2\right)+\frac{1}{4L^4\pi^4 T_0^4}\right)\,.
\ee
The flat space limit can be found by taking the $L\pi T_0\to \infty$ limit of $\varepsilon$ while making $h^{\mu\nu}\to \eta^{\mu\nu}$.

\section{Single nonzero  chemical potential}\label{app:singlechem}

For a single charge or chemical potential $\mu_1=\mu$, $Q_1=Q$, $\mu_2=\mu_3=0$, $Q_2=Q_3=0$ we can set normalized charge variables to be $x_1=x$, $x_2=x_3=0$. Then, the grand canonical potential is
\be
\Omega_1 =- N_c^2\frac{\pi^2}{8}T_0^4\left(1+x^2 -\frac{1}{L^2\pi^2 T_0^2}\right) \,.
\ee

 \begin{figure}[t!]
   \begin{center}
   \includegraphics[scale=0.8]{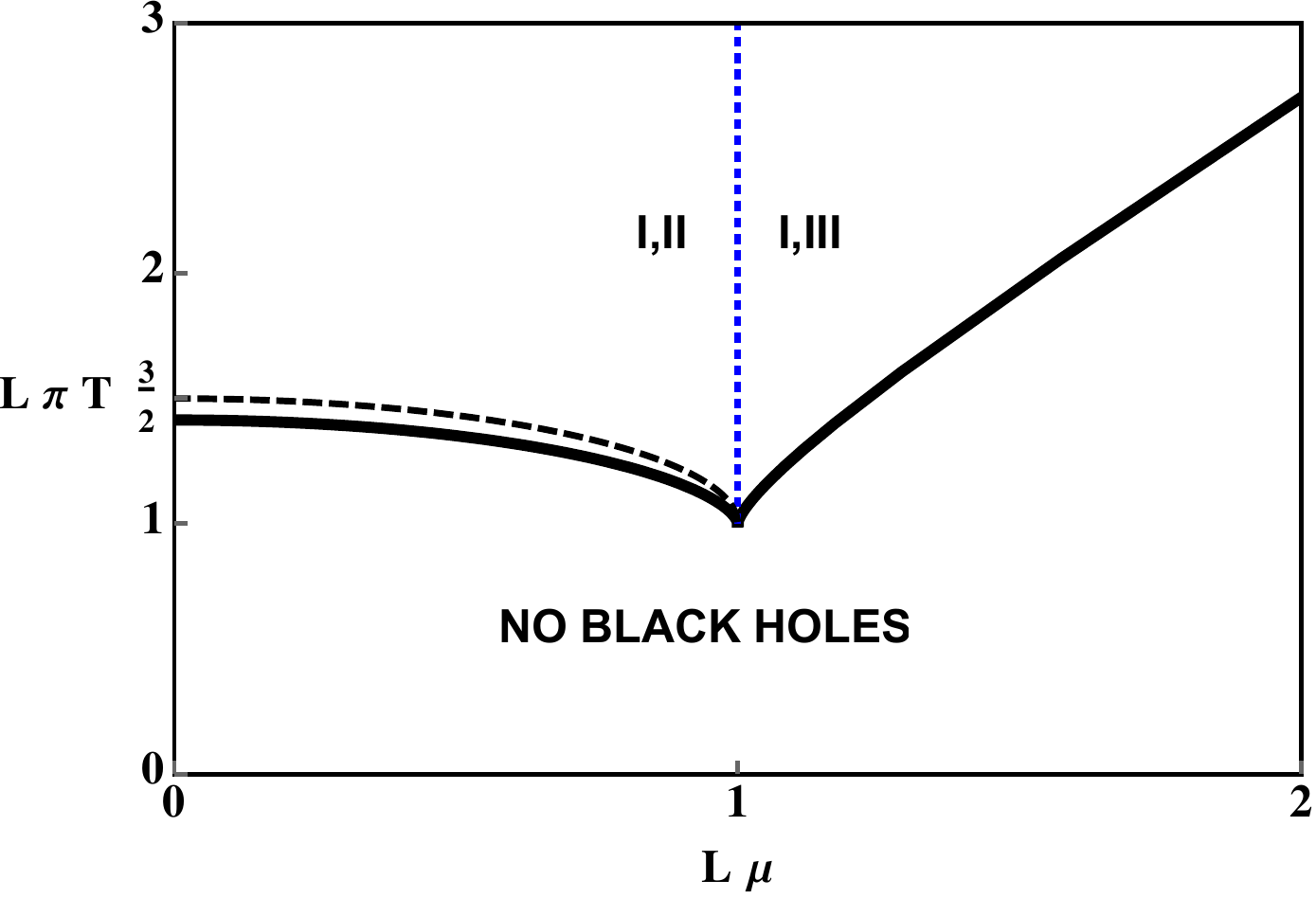}
     \caption{\small Phase diagram in the grand canonical ensemble for a single nonzero chemical potential. Susceptibilities diverge at the black thick curves, and black hole solutions exist only in the region between the curves. The blue dashed line at $L\mu=1$ separates the unstable branches {\bf{II}} and {\bf{III}}. The black dashed line marks the location of the Hawking-Page transition. Notice that there is no nucleation instability in this case.}\label{fig:gcphd1}
   \end{center}
   \end{figure}

\begin{figure}[t!]
   \begin{center}
\begin{tabular}{ccc}
   \includegraphics[width=\textwidth/4]{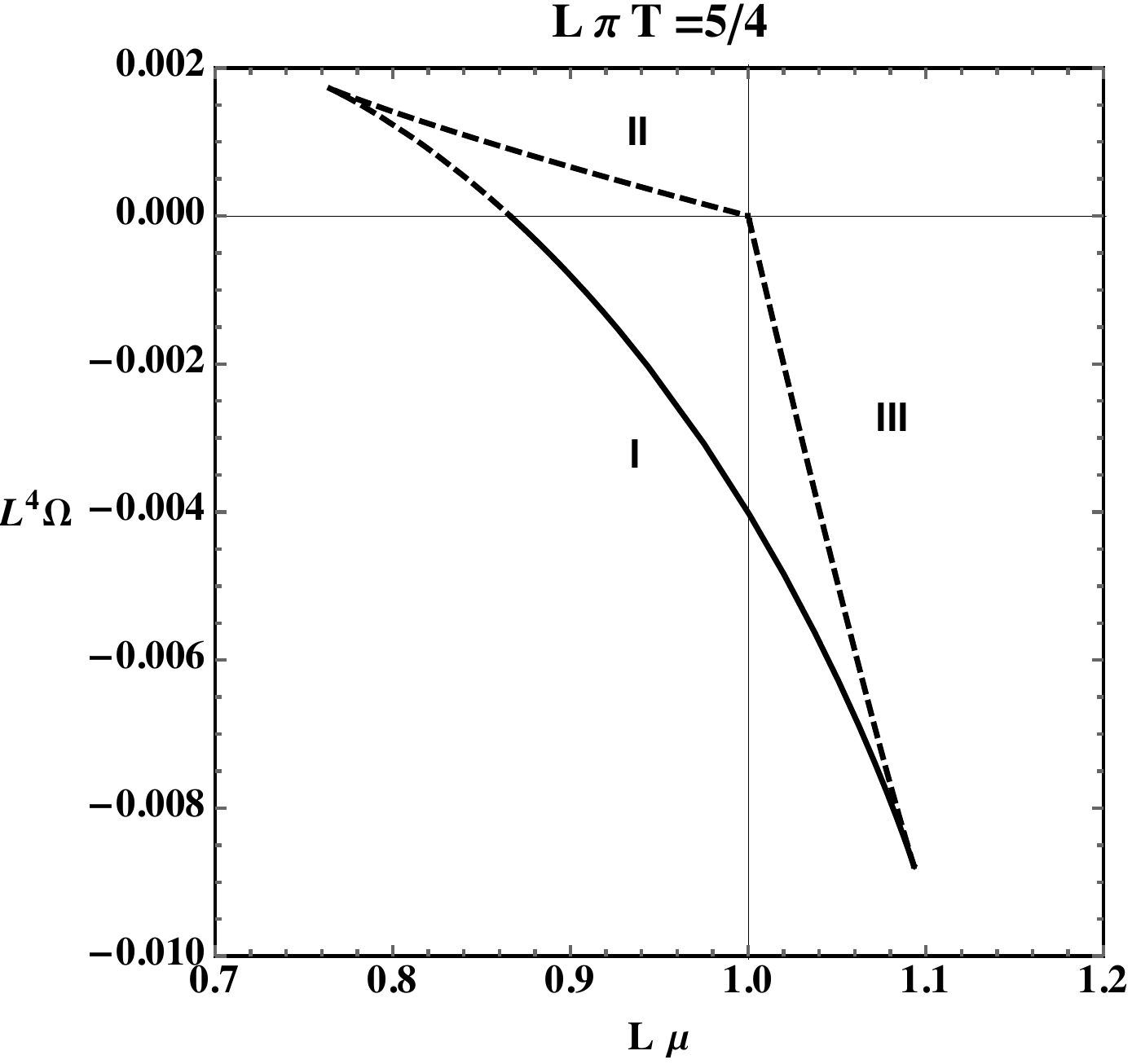} &
   \includegraphics[width=\textwidth/4]{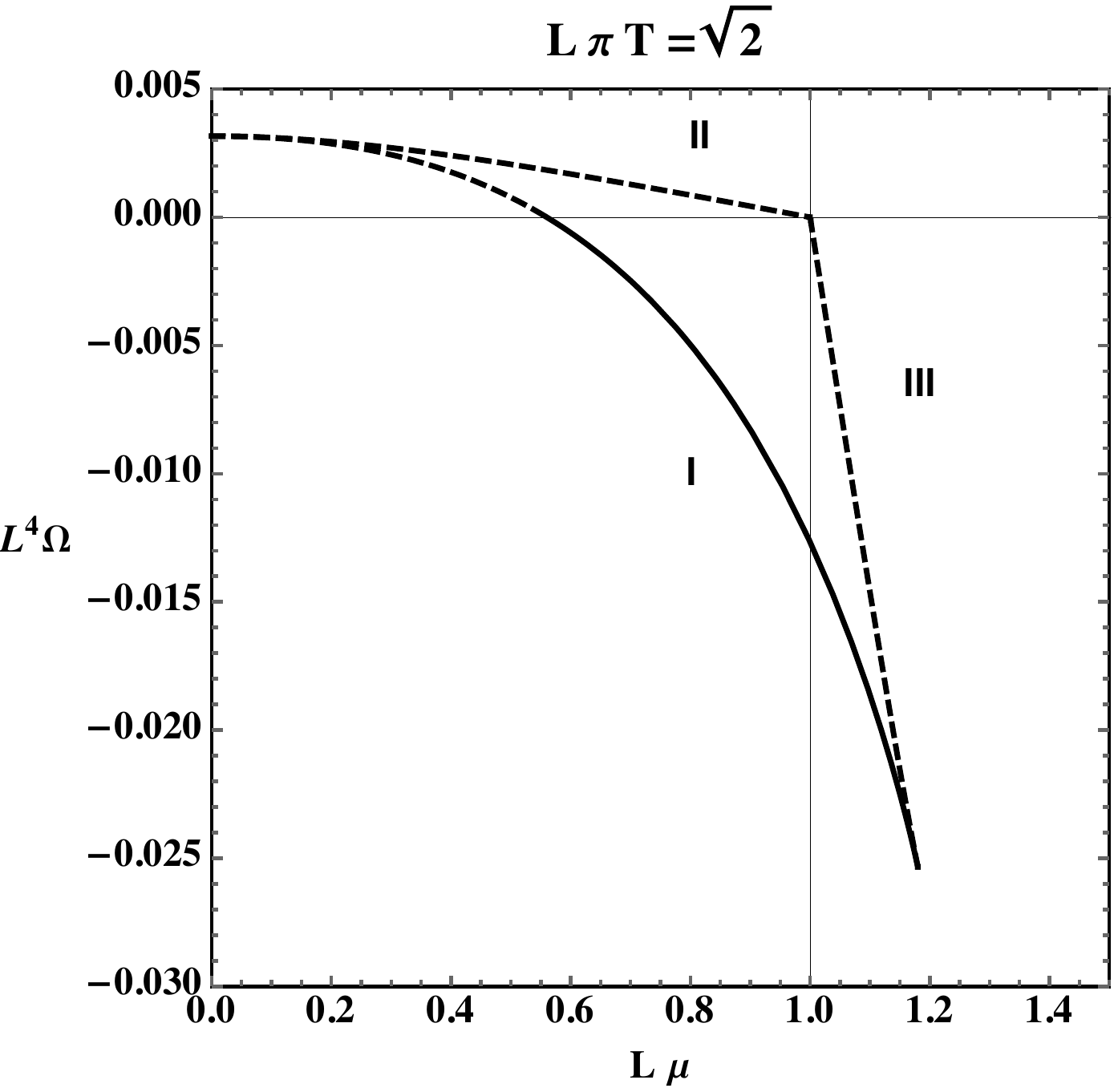} &
   \includegraphics[width=\textwidth/4]{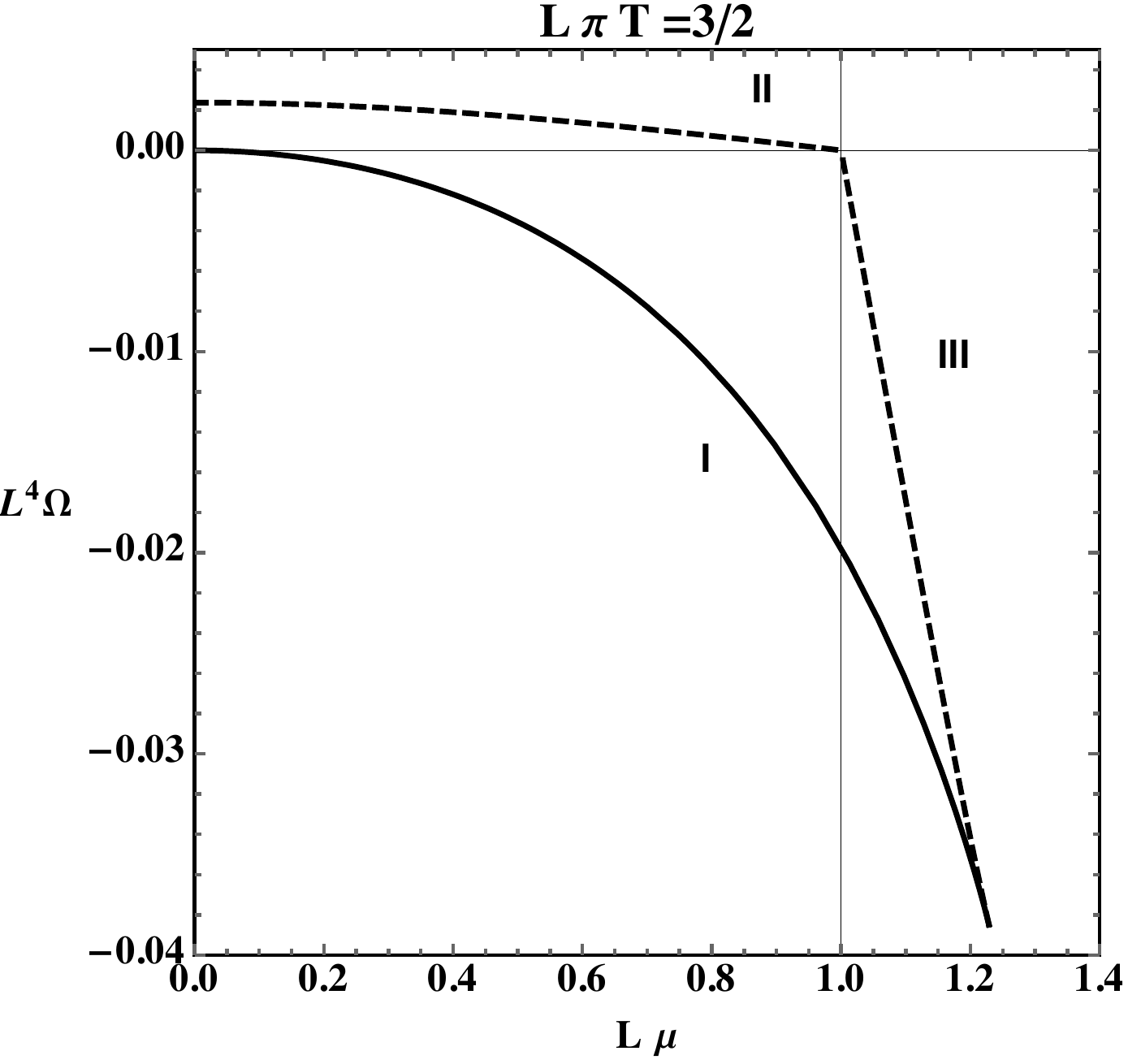}
   \end{tabular}
     \caption{\small Grand canonical potential as function of the chemical potential for different values of the temperature. Dashed curves correspond to unstable or metastable phases. The branches with larger values of $\Omega$ correspond to the unstable phases, {\bf{II}} and {\bf{III}}, while the lower branch ({\bf{I}}) is metastable until it reaches $\Omega=0$, which corresponds to the Hawking-Page transition. Susceptibilities diverge at the point where the two symmetric branches touch. At high enough temperatures, $L\pi T>3/2$, the Hawking-Page transition goes away and the stable  black hole phase dominates at low values of the chemical potential. In all cases, at large enough values of the chemical potential, the black hole phases disappear.}\label{fig:gcpot1}
   \end{center}
   \end{figure}
 
In the grand canonical ensemble black hole solutions exist in a region of the $(L\mu,L\pi T)$ plane limited by the black curves in Fig.~\ref{fig:gcphd1}, which are determined by the conditions
\be
\frac{1}{t_0^2}  = \frac{1}{2} \left(1+2 x^2\pm\sqrt{8 x^2+9}\right)\,. \label{eq:curvecond}
\ee
One curve (with a plus sign in \eqref{eq:curvecond})  interpolates between $(L\pi T,L\mu)_+=(\sqrt{2},0)$ for $x\to 0$ and  $(L\pi T,L\mu)_+=(1,1)$ for $x\to\infty$. The other curve is defined for $x>\sqrt{2}$ and starts at $(L\pi T,L\mu)_-=(1,1)$ when $x\to\infty$ and approaches a straight line $L\mu=L\pi T/\sqrt{2}$ in the limit $x\to \sqrt{2}$. At these curves susceptibilities diverge and black hole solutions do not exist outside the region delimited by the two curves. Therefore, planar black hole solutions exist only for $\mu < \frac{1}{\sqrt{2}}\pi T$. In the region where solutions exist there are three branches, one thermodynamically stable ({\bf{I}}) and the other two unstable ({\bf{II}} and {\bf{III}}). Each of the unstable branches exists only for either $L\mu<1$ ({\bf{II}}) or $L\mu >1$ ({\bf{III}}), while the stable branch covers the whole allowed region. The value of the grand canonical potential for each branch is represented in Fig.~\ref{fig:gcpot1}, with dashed curves corresponding to unstable or metastable phases.  As in the case of three equal chemical potentials, the Hawking-Page transition is localized away from the limiting curves and for $L\pi T>3/2$ the stable solution is the dominant phase at small values of the chemical potential. For large enough values of the chemical potential the stable and unstable branches merge and black hole solutions stop existing. Finally, we note that in the entire region where black hole solutions exist, the brane nucleation instability is suppressed since typical branes in the stack do not have enough angular momentum to nucleate in the bulk. In the region where one would have expected it to be unsuppressed, {\emph{i.e.}}, for large $\mu/T$ there are no black hole solutions, see Fig.~\ref{fig:gcphd1}.

\bibliographystyle{JHEP}
\bibliography{refs}

\end{document}